\documentclass[modern]{aastex631}
\usepackage{amsmath,array,graphicx}
\usepackage{tabularx}

\def\lapprox{\hbox{\lower .8ex\hbox{$\,\buildrel < \over\sim\,$}}}
\def\gapprox{\hbox{\lower .8ex\hboxeening{$\,\buildrel > \over\sim\,$}}}

\def\kmso{\hbox{${\rm km}\:{\rm s}^{-1}$}}
\def\kms{\hbox{${\rm km}\:{\rm s}^{-1}\,$}}

\def\aap{\hbox{A\&A}}

\newcommand{\RomanNumeralCaps}[1]
 {\MakeUppercase{\romannumeral #1}}

\begin{document}

\title
{A possible surviving companion of the SN Ia in the Galactic SNR G272.2-3.2}

\author{P. Ruiz-Lapuente}
\email{pilar@icc.ub.edu}
\affiliation{Instituto de F\'{\i}sica Fundamental, Consejo Superior de 
Investigaciones Cient\'{\i}ficas, c/. Serrano 121, E-28006, Madrid, Spain}
\affiliation{Institut de Ci\`encies del Cosmos (UB--IEEC),  c/. Mart\'{\i}
i Franqu\'es 1, E--08028, Barcelona, Spain}

\author{J.I. Gonz\'alez Hern\'andez}
\affiliation{Instituto de Astrof\'{\i}sica de Canarias, E-38200 La Laguna, 
Tenerife, Spain}
\affiliation{Universidad de La Laguna, Dept. Astrof\'{\i}sica, E38206 
La Laguna, Tenerife, Spain}

\author{R. Cartier}
\affiliation{Gemini Observatory, {\it NSF}'s National
Optical-Infrared Research Laboratory, Casilla 603, La Serena, Chile}

\author{K. Boutsia}
\affiliation{Las Campanas Observatory, Carnegie Institution for Science,
  Colina El Pino, Casilla 601
La Serena, Chile}

\author{F. Figueras}
\affiliation{Departament de F\'{\i}sica Qu\`antica i Astrof\'{\i}sica, 
Universitat de Barcelona, c/. Mart\'{\i} i Franqu\'es 1, E--08028 Barcelona, 
Spain}
\affiliation{Institut de Ci\`encies del Cosmos (UB--IEEC),  c/. Mart\'{\i}
i Franqu\'es 1, E--08028, Barcelona, Spain}
\affiliation{Institut d'Estudis Espacials de Catalunya (IEEC), E-08438, 
Barcelona, Spain}

\author{R. Canal}
\affiliation{Departament de F\'{\i}sica Qu\`antica i Astrof\'{\i}sica, 
Universitat de Barcelona, c/. Mart\'{\i} i Franqu\'es 1, E--08028 Barcelona, 
Spain}
\affiliation{Institut d'Estudis Espacials de Catalunya (IEEC), E-08438, 
Barcelona, Spain}

\author{L. Galbany}

\affiliation{Institute of Space Sciences (ICE, CSIC), Campus de Bellaterra, 
Can Magrans s/n, E-0893, Bellaterra, Spain}

\affiliation{Institut d'Estudis Espacials de Catalunya (IEEC), E-08438, 
Barcelona, Spain}

\begin{abstract}

  \noindent
  We use the {\it Gaia} EDR3 to explore the Galactic supernova remnant SNR G272.2-3.2, produced by the explosion of a Type Ia supernova (SNIa), about 7,500 years ago, to search for a surviving companion. From the abundances in the SNR ejecta, G272.2-3.2 is a normal SN Ia. The {\it Gaia} parallaxes allow to select the stars located within the estimated distance range of the SNR, and the {\it Gaia} proper motions to study their kinematics. From the {\it Gaia} EDR3 photometry, we construct the HR diagram of the selected sample, which we compare with the theoretical predictions for the evolution of possible star companions of SNIa. We can discard several proposed types of companions by combining kinematics and photometry. We can also discard hypervelocity stars. We focus our study on the kinematically most peculiar star, {\it Gaia} EDR3 5323900215411075328 (hereafter MV-G272), a 8.9 $\sigma$ outlier in proper motion. It is of M1-M2 stellar type.
Its trajectory on the sky locates it at the center of the SNR, 6,000--8,000 years ago, a unique characteristic among the the sample. Spectra allow a stellar parameters determination and a chemical abundance analysis. In conclusion, we have a candidate to be the surviving companion of the SN Ia that resulted in SNR G272.2-3.2. It is supported by its kinematical characteristics and its trajectory within the SNR. This opens the possibility of a single-degenerate scenario for a SN Ia with an M-type dwarf companion.

\end{abstract}


\section{Introduction}

Type Ia supernovae (SNe Ia) are powerful calibrated candles, whose use as 
distance indicators in 
cosmology led to the discovery of the accelerated expansion of the universe
(Riess et al. 1998; Perlmutter et al. 1999), and currently are major tools 
in the exploration of the nature of dark energy (Rose et al. 2020; Hayden et 
al. 2021). Besides, SNe Ia are the main producers of the Fe--peak elements 
in the universe (see, for instance, Branch \& Wheeler 2017).

\noindent
However, there is still a lack of knowledge concerning 
the exact nature of the progenitors of the SNeIa: on their explosion mechanism 
and on 
the kind of stellar systems from which they arise, both points being closely 
related. There is now a universal agreement that they are produced by the 
thermonuclear explosion of a white dwarf made of carbon and oxygen (a C+O WD),
with a mass not far from the Chandrasekhar mass. But the explosion might be 
initiated close to the center of the star, when the mass reaches the 
Chandrasekhar limit due to accretion of material from a close binary companion 
(Whelan \& Iben 1973),
or result from compression produced by the detonation of a helium layer close
to the surface of the WD (Livne 1990; Livne \& Arnett 1995) or by the 
collision with another WD (Rosswog et al. 2009). In the last two
cases, the exploding WD would have a mass below the Chandrasekhar mass. 

\noindent
The WD progenitor of the SN Ia must be in a close binary system in all cases.
The companion,
the mass donor, may either be a star in any 
stage of thermonuclear burning (Whelan \& Iben 1973; Nomoto 1982), that being 
called the single--degenerate (SD) scenario, or be another WD (Webbink 1984;
Iben
\& Tutukov 1984): the double--degenerate (DD) scenario. The core--degenerate
(CD) model of SN Ia explosion (Kashi \& Soker 2011; Soker 2013), in 
which a WD merges with the electron--degenerate core of an asymptotic 
giant--branch (AGB) star can be included within the DD scenario.  

\noindent
It is unknown what fraction of the observed SNeIa corresponds 
to each of the two scenarios. In the DD channel, in most cases considered, both
WDs should be destroyed  by the explosion, no bound remnant being left. There
is a possible exception, though, in the case of explosions triggered by
the detonation of a surface layer made of He, accreted by the exploding WD
from a less massive WD companion: then, the outburst might happen when the
mass--donor has not yet been tidally disrupted. Due to its very high orbital
velocity, the WD companion should be ejected as a {\it hypervelocity star} 
($ v > 1,000$ ~\kmso). This happens in the {\it dynamically driven 
double--degenerate, double--detonation} scenario, D$^{6}$ (Shen \& Moore 2014;
Shen \& Schwab  2017; Shen et al. 2018).

\noindent
In
the SD case, the binary companion of the exploding WD survives (Marietta et al.
2000; Pakmor et al. 2008; Pan et al. 2012a,b). The surviving companion might 
be in any evolutionary stage: main sequence, subgiant or red giant, be a 
helium  or a sdB star (see Wang \& Han 2012; Maoz et al. 2014; Ruiz--Lapuente 
2014, 2019, for reviews). Of course, the detection of such companions at the
location of a SN Ia would confirm the SD scenario (for that particular SN at
least and thus for some fraction of them).

\noindent
In the hydrodynamical simulations of the impact of the SN Ia ejecta with the
companion, different kinds of stars have been considered: main--sequence (MS)
stars (Marietta et al. 2000; Pakmor et al. 2000; Pan et al. 2012a;
McCutcheon et al. 2022), subgiants
(SG) (Marietta et al. 2000; Pan et al. 2012a), helium stars (Pan et al. 2012a;
 Liu et al. 2013a),
red--giant (RG) stars (Marietta et al. 2000; Pan et al. 2012a), and sdB stars 
(Bauer et al. 2019). These
calculations predict the state of the companion just after the SN Ia explosion.
Different amounts of mass have been stripped by the impact with the ejecta and
the stars are bloated and overheated. Those results provide
 the initial conditions to calculate the subsequent evolution of the
companion stars.

\noindent
The time evolution of possible SN Ia companions, on scales from hundreds to
thousands of years, has been calculated by Podsiadlowski (2003), for a SG 
companion, Pan et al. (2012b, 2014), Shappee et al. (2013) and  Rau \& Pan 
(2022), for MS and SG 
companions, by Pan et al. (2012b, 2014) for RGs, by Bauer et al. (2019) 
for sdB stars,  and by Liu et al. (2022) for He stars. The calculations 
predict
the changes in luminosity and effective temperature of the stars, starting  
from the time they recover hydrostatic equilibrium after experiencing the 
impact of the SN ejecta. The stars are then overluminous as compared with 
their previous state and they evolve, on thermal time scales, to meet the
characteristics corresponding to the new mass and thermonuclear burning 
stage. Pan et al. (2014) have also calculated the chemical pollution 
of the atmospheres by the ejecta.  

\noindent
One effect due to the star having been in a close binary system previous
to the explosion are high space velocities (due to their orbital velocities 
prior to the disruption of the binary, in addition to the kick imparted by the 
ejecta)\footnote{Companions that orbit at large separations from the
exploding white dwarf would have lower space velocities when the system is
disrupted. Some red giants companions would have low peculiar velocities,
but those surviving stars would be too luminous companions (Marietta, Burrows \&
Fryxell 2000) and have not been seen in the SN Ia remnants explored so far.}. 
Therefore, when searching for possible 
companions within the remnants of recent SNe Ia, one should look for high 
spatial 
velocities, paying attention to the past trajectories, and also for anomalous 
positions 
in the color--magnitude and color--color diagrams of the stars close to the 
center of the SNR. We should also look for possible
chemical enrichment, in Fe--peak elements namely. Given the current
observational means, only remnants of SNeIa that took place in our
Galaxy or in the Large Magellanic Cloud (LMC) have been explored in search for
surviving companions, at present. 

\noindent
Up to 14 supernova remnants (SNRs) of the Ia type have been identified in the
Galaxy and 12 in the LMC (Ruiz--Lapuente 2019). Of the former, only three have
been explored (corresponding to the ``historical'' SNe Ia). They are those of
SN 1572, or Tycho Brahe's SN (Ruiz--Lapuente et al. 2004, 2019; Gonz\'alez
Hern\'andez et al. 2009; Kerzendorf et al. 2009, 2013, 2018a; Bedin et al.
2014), of SN 1604, or Kepler's SN (Kerzendorf et al. 2014; Ruiz--Lapuente et 
al. 2018), and of SN 1006 (Gonz\'alez Hern\'andez et al. 2012;
Kerzendorf et al. 2012, 2018b; Shields et al. 2022). No indisputable 
 companion candidate has
been found ins any of them. In the case of SN 1006, the absence of candidates
points to a DD origin of the SN. In that of SN 1604 the same absence,
joined to the characteristics of the SNR suggested the CD scenario
(Ruiz--Lapuente et al. 2018).  
A candidate has been found for SN 1572, but the identification is in dispute 
(see the references above).

\noindent
Five SNRs of the Ia type have been explored in the LMC: SNR 0509--67.5
(Schaefer \& Pagnotta 2012; Litke et al. 2017), SNR 0519--69.0 (Edwards 
et al.
2012; Li et al. 2019), SNR N103B (Li et al. 2017), SNR 0505--67.9 (DEML71), and
SNR 0548--70.4 (Li et al. 2019). No clear surviving companion candidate has
been found in any of them but a star in N103B has characteristics compatible
with being a surviving SG (Li et al. 2017), and two other stars, in 0519--69.0
and DEML71 respectively, have large radial velocities and they might also be
SN companions (Li et al. 2019). 

\noindent
Out of the still unexplored SN Ia Galactic SNRs, most are at large distances 
and located close to the Galactic plane, which causes them to be very heavily 
reddened.  
 
\noindent   
That is not the case, however, of SNR G272.2--3.2, at a distance $\sim$ 1--3
kpc. The EDR3 of {\it Gaia} now provides the parallaxes, proper motions and
photometry allowing a first exploration of the central region of this SNR.
Knowledge of the parallaxes allows to select the
stars, close to the center of the SNR, which are at distances compatible with 
that of the remnant. We look for peculiar proper motions and compare
the HR diagram of the sampled stars with the evolutionary paths 
predicted for different types of surviving companions. That already allows us to
exclude the presence of several kinds of proposed candidates and to 
select stars deserving further analysis. 

\noindent
In the next Section we summarize the characteristics of the SNR G272.2--3.2
and 
we define the search area for the possible companion star.
Observations are described in Section 3. 
Proper motions, their transformation to tangential velocities and a kinematic 
outlier are treated in Section 4.
Reddening of the observed field is discussed together with the stellar spectra
obtained, in Section 5. The stellar parameters of our unique 
peculiar star are determined and a chemical analysis is done in Section 6. In
Section 7, the characteristics of this star are further discussed.   
Color--magnitude  and the HR
diagrams, along with their comparison with the evolutionary tracks predicted
for different types of companions are dealt with in Section 8. Exploration 
of more extended areas than in Section 2 and search for the possible
presence of hypervelocity stars are examined in Section 9. All results are
summarized and conclusions drawn in the final Section.   
      
\begin{figure}[ht!]
\centering 
\includegraphics[width=0.50\textwidth]{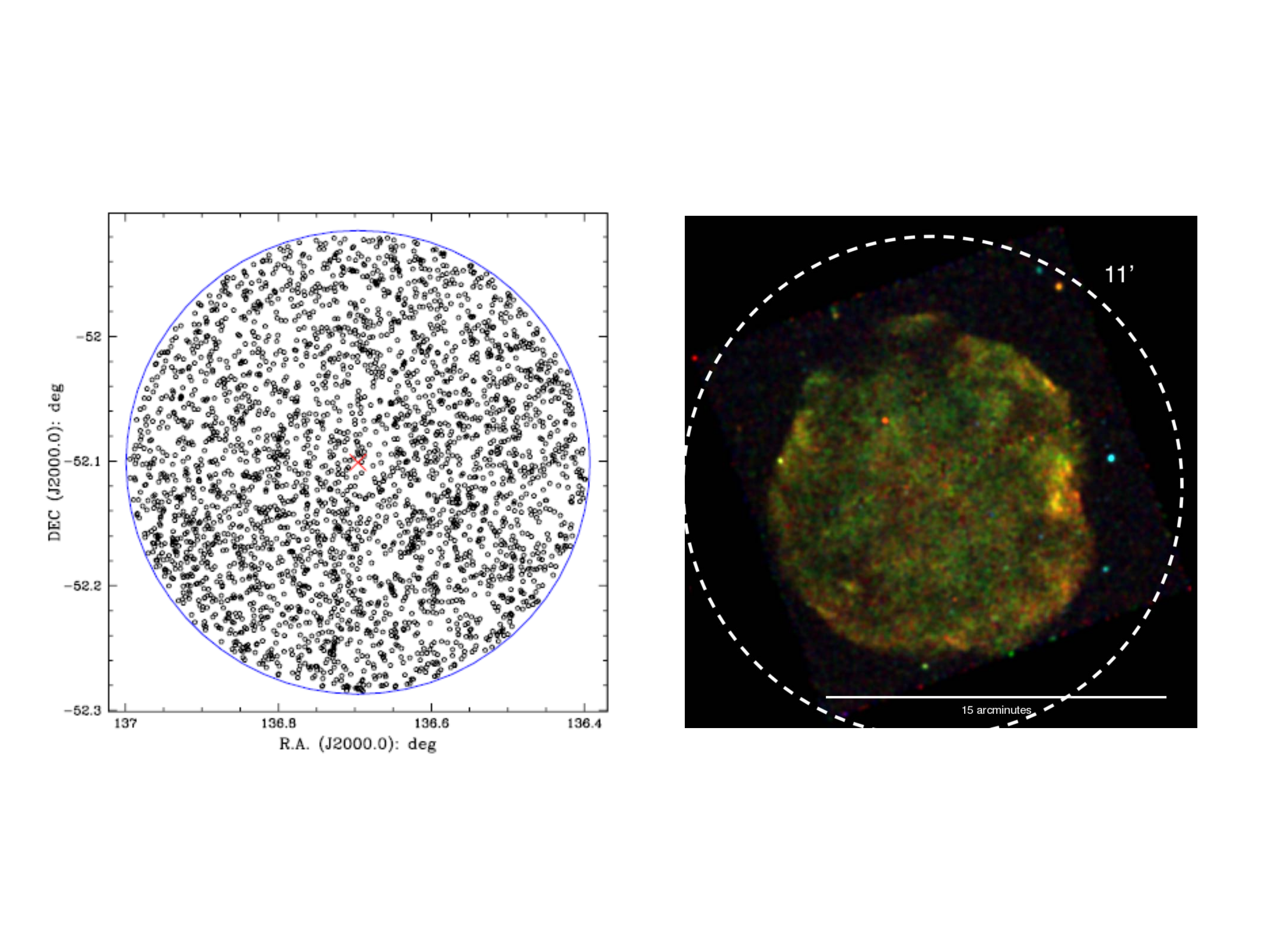}
\caption{Left panel: positions of the 3,082 stars in our sample. The red
  cross marks the
  centroid of the G272.2-3.2 SNR, and the blue circle corresponds to the 11
  arcmin radius around it. Right panel: the 11 arcmin radius circle 
  superimposed to an X--ray image of the SNR.} 
\label{Figure 1}
\end{figure}

\section{G272.2--3.2}    
\noindent
The SNR G272.2--3.2 was discovered in X--rays by Greiner \& Egger (1993) 
during 
the {\it ROSAT All Sky Survey}, details being given in Greiner, Egger \&
Aschenbach (1994). Radio observations by Duncan et al. (1997) measured a 
diameter $\sim$15 arcmin for the remnant. It has later been studied, in
X-rays, by    
Harrus et al. (2001), Lopez et al. (2011),   
McEntaffer et al. (2013), Yamaguchi et al. (2014) and Kmitsukasa et al. (2016),
with mounting evidence, from the measurement of overabundances of Ar, Ca, 
Si, S, Fe and Ni, that it was produced in a SNIa. 

\noindent
{\it Chandra} observations
have provided measurements of chemical abundance ratios that are in good 
agreement with the predictions for delayed detonation models of SN Ia 
explosions (Sezer \& G\"ok 2012).

\noindent
The SNR is at a  distance  $d = 1.8^{+1.4}_{-0.8}$
kpc (Greiner et al 1994) or $d \sim 2-2.5$ kpc (Harrus et al. 2001;
Kamitsukasa et al. 2016). Its age is estimated to be 7,500$^{+3800}_{-3300}$ yr 
(Leahy et al. 2020; Xiang \& Jiang 2021). 

\noindent
It is located about 110 pc below the Galactic plane. 
 The radius of the remnant is about 8 arcmin, 
and its centroid
lies at $\alpha_{\rm J200}$ = 09$^{h}$ 
06$^{m}$ 45$^{s}$.7, $\delta_{\rm J2000}$ = -52$^{o}$ 07' 03'' (Greiner \& Egger
1993), which corresponds to the Galactic coordinates $l$ = 272$^{o}$ 12' 36.9'',
$b$ = -3$^{o}$ 10' 34.4''.

\noindent
We have searched in the {\it Gaia} EDR3 database for the stars within a radius
of 11 arcminutes (thus extending beyond the whole SNR), and with parallaxes 
corresponding to distances 1 kpc $\leq d \leq$ 3 kpc. That has produced a 
sample of 3,082 stars (see Figure 1). The 11 arcmin radius is slightly above 
the arc described by a star moving at 500 ~\kmso, perpendicularly to the line 
of sight and a distance of 2 kpc, in 12,000 yr. Wider search radii 
are considered and the corresponding results presented and discussed in 
Section 9.

Besides parallaxes and proper motions, we have also extracted the {\it Gaia}
photometry of the stars in our sample (left panel of Figure 2). The 
photometry in Figure 2 is still
uncorrected for interstellar extinction and reddening.

\begin{figure}[ht!]
\centering
\includegraphics[width=0.45\textwidth]{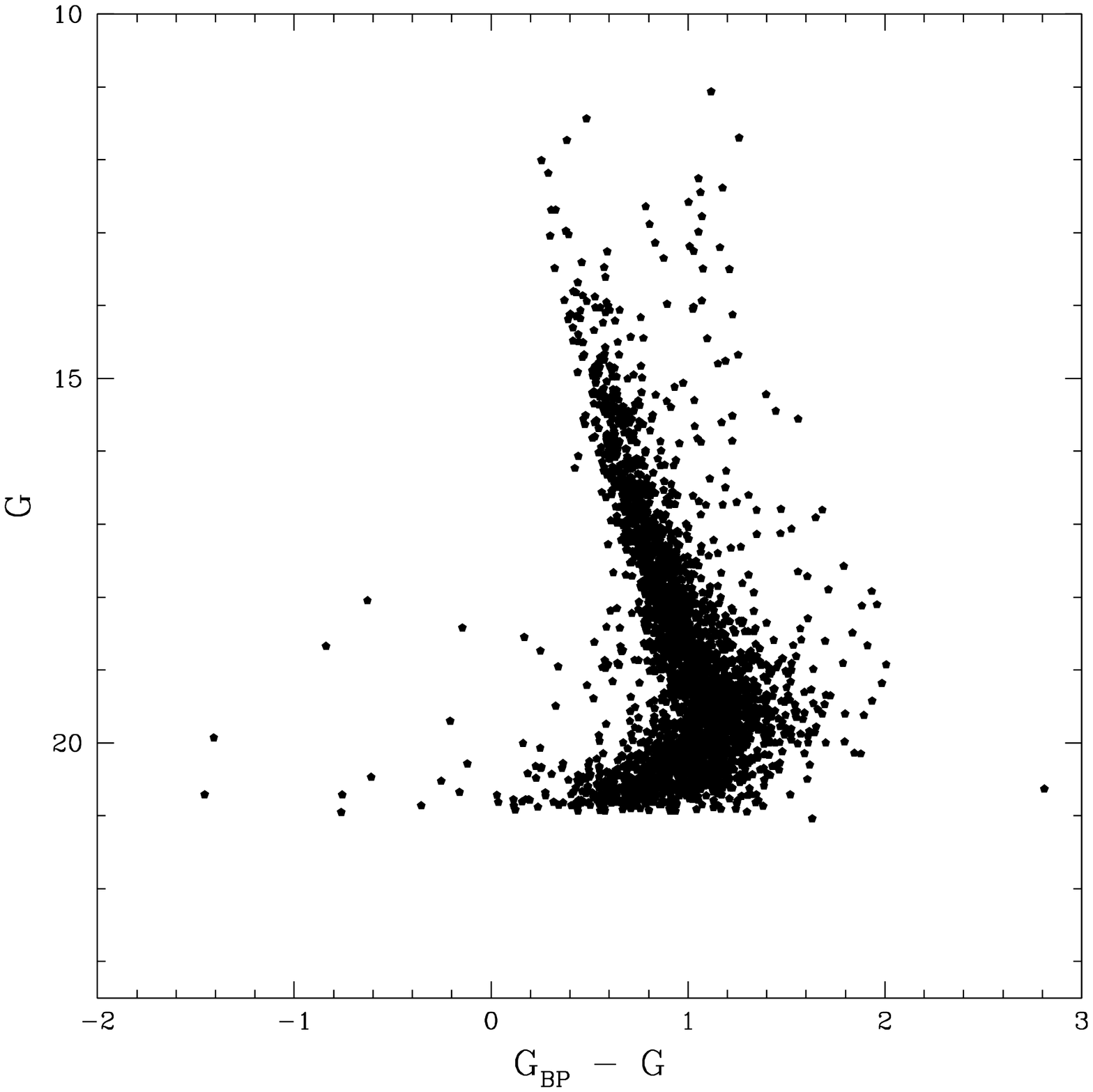}
\includegraphics[width=0.45\textwidth]{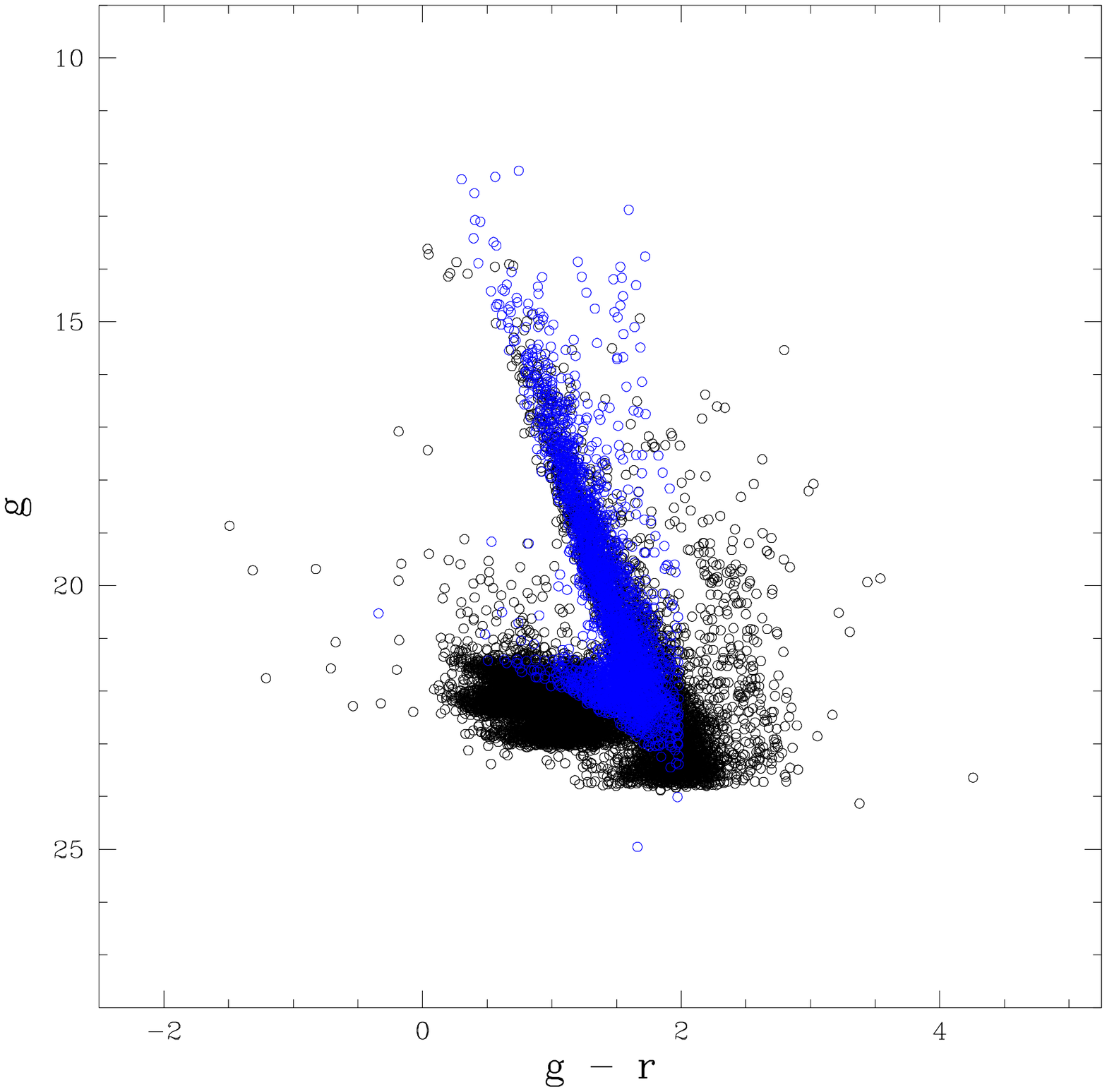}
\caption{Left: The $G$ $vs$ $G_{BP} - G$ diagram for the stars in our
 sample, uncorrected from reddening and extinction. Right: A $g$  $vs$ $g - r$ 
diagram showing the stars of our sample (in blue) superimposed 
to those  obtained from the {\it DECaPS} survey (in black), 
within the same cone but with no
  limitation on distances (which are unknown there). We see the
consistency between the {Gaia} EDR3 photometry and that of the {\it DECaPS}  
survey  (see main text). 
The {\it Gaia} magnitudes have been transformed into the {\it SDSS} magnitudes
using the expressions given by Carrasco ({\it Gaia Data Release Documentation} 
5.3.7). Errors of up to 0.16 mag can be made when transforming $G$ into 
$g$ magnitudes.}

\label{Figure 2}
\end{figure}

\bigskip
\bigskip
\bigskip
\bigskip

\section{Observations}

\subsection{Gaia EDR3}

\noindent
As stated above, the {\it Gaia} EDR3 has been used to obtain the proper
motions, parallaxes and photometry in the $G$, $G_{BP}$ and $G_{RP}$ bands, for
the stars within a circle of 11 arcmin radius on the sky, around the centroid
of SNR G272.2--3.2, and parallaxes corresponding to distances in the 
selected range.

\noindent
{\it Gaia} EDR3, released on December 2020, contains the full astrometric 
solution (positions, parallaxes and proper motions) for around 
1.468$\times10^{9}$ 
stars, with a limiting magnitude $G \sim$ 21 mag. 

\noindent
It also gives $G$ magnitudes
for 1.806$\times10^{9}$ sources, and $G_{BP}$ and $G_{RP}$ for around 
1.542$\times10^{9}$ and 1.555$\times10^{9}$ sources, respectively. It is 
publicly available\footnote{gea.esac.esa.int/archive/}.
\subsection{DECaPS}

\noindent
We have also used the photometric data from {\it DECaPS} ({\it DECam Plane
  Survey}). This is a five--band optical and infrared survey of the southern
Galactic plane with the {\it Dark Energy Camera} ({\it DECam}) at Cerro Tololo.
It covers about 1,000 square degrees (the low latitude Galactic plane south
of $\delta < 30^{o}$).

\noindent
The survey, which is publicly available\footnote{http://decaps.skymaps.info}, 
has a depth of 23.7, 22.8, 22.2, 21.8 and 21.0 mag in the 
{\it grizY} bands. We have explored the same circle of 11 arcmin radius around
the centroid of G272.2--3.2 as in {\it Gaia} EDR3, but with no limits on 
distance here (parallaxes unknown). That has yielded 38,019 
stars with complete {\it gri} photometry at least. Our {\it Gaia} DR3 sample
sits at the core of this more extended sample, which shows the consistence
between the two photometric systems (see the right panel of Figure 2).
 There, {\it Gaia} magnitudes have been transformed into the {\it SDSS} 
magnitudes using the expressions given by Carrasco 
({\it Gaia Data Release Documentation} 5.3.7). An error by up to 0.16 mag can 
be made when going from $G$ to $g$ magnitudes).

\subsection{Spectra}

\noindent
Several stars have been
observed using the camera on the {\it Goodman} spectrograph (Clemens et al.
2004), mounted on the 4.1m {\it SOAR} telescope. The 600 lines mm$^{-1}$ grating
and the 1.0 arcsec slit have been used, providing a resolution of
$\sim$4.3 \AA \ or better (R$\sim$1,400)
and covering from 4,550 \AA  to 7,050 \AA . We reduced the {\it Goodman} data 
following usual
steps, including bias substraction, flat--fielding, cosmic ray rejection
using {\it LA--Cosmic} (van
Dokkum 2001), wavelength calibration, flux calibration, and telluric correction
using our own custom {\it IRAF} routines. The telluric correction was performed using a flux standard observed at the beginning of the night with the same configuration of our science targets. A final combined spectrum was obtained by combining the individual spectra, weighted by the signal-to-noise ratio.

\noindent
Spectra of the same stars have also been obtained with the {\it MIKE}
spectrograph\footnote{Bernstein, R., Shectman, S.A., Gunnels, S.M., et al. 
SPIE, 4841, doi:0.117/12.461502} at the 6.5m {\it Clay} telescope. {\it MIKE} 
is a high-resolution
optical spectrograph with a wavelength coverage from 3,500 to 9,500 \AA.
The stars have been observed several nights (see Table 1). We used the Red 
arm, covering 
the wavelength range
4,000-9,500 \AA, with a slit width of 0.7 arcsec and a binning of 2$\times$2 
providing a resolving power of R$\sim$28,000 (equivalent to a FWHM $\sim$ 10.7 
~\kms and a pixel size of 0.069 \AA \ (equal to 2.69 ~\kms). 

\noindent
The {\it CarPy} pipeline (Kelson et al. 2000; Kelson 2003) has been used to
reduce each single
night separately. The final product was a sky-substracted and
wavelength-calibrated spectrum for each separate order per night. A
spectrophotometric standard star has also been observed each night in
order to perform relative flux calibration. This was done using {\it IRAF}
routines ({\it standard}, {\it sensfunc} and {\it calibrate}), resulting in a
spectrum calibrated to the correct flux scale and corrected for extinction. In
order to obtain the final stacked spectrum, the flux-calibrated spectrum for
each order in each night has been combined using the {\it IRAF} task
{\it scombine}. This led to an 1D spectrum over the full wavelength range.

\begin{table*}
\caption{\ \ Summary of Spectroscopic Observations}
\begin{center}
\begin{tabular}{c c c c c }
\hline
\hline 
\\
 &    & {\it SOAR} telescope/{\it Goodman} spectrograph & &   \\ 
\hline
Date  & Source & exp.time & slit  & airmass   \\
\hline 
Feb07 &  {\it Gaia} EDR3 5323900211541075328  & 2hr & 1.0   & 1.20   \\
Mar10 & {\it Gaia}  EDR3 5323871314998012928  & 1hr45min & 1.0  & 1.10   \\
Apr26 & {\it Gaia}  EDR3 5323852210990643584  & 1h45min & 1.0  &  1.20    \\
\hline
\\
 &  &  {\it Clay} telescope/{\it MIKE} spectrograph & &   \\ 
\hline
Nov20 &  {\it Gaia} EDR3 5323900211541075328 & 1hr      & 0.70  & 1.20   \\
Nov21 &  {\it Gaia} EDR3 5323900211541075328 & 1hr20min & 0.70   & 1.15 \\
Feb25 &  {\it Gaia} EDR3 5323900211541075328 & 1hr20min & 0.70  & 1.12  \\
May19 &  {\it Gaia} EDR3 5323900211541075328 & 2hr      & 0.70  & 1.23  \\
May21 & {\it Gaia}  EDR3 5323871314998012928 & 2hr      & 0.70  & 1.23 \\
\hline
\hline
\end{tabular}
\end{center}
\end{table*}

\section{Space velocities.}

Peculiar space velocities as compared with those
of the surrounding stars are among the likely characteristics of surviving 
companions of SNe Ia explosions. In Figure 3 we show the distribution of 
proper motions in right ascension, RA (left) and in declination, DEC (right), 
The mean proper motion in RA ($\mu^{*}_{\alpha}$) is -4.8 mas/yr, with a 
standard deviation $\sigma$
= 3.12 mas/yr, while in DEC ($\mu_{\delta}$) we have a mean of 4.44 mas/yr with 
$\sigma$ = 3.11
mas/yr. There is a star,  {\it Gaia} EDR3 5323900211541075328  
(M\RomanNumeralCaps{5}-G272)
(RA = 09$^{h}$ 06$^{m}$ 24.66$^{s}$; DEC = -52$^{o}$ 03' 09.684'', $G$ = 
19.854 mag) with 
$\mu^{*}_{\alpha}$ = -22.80 mas/yr and $\mu_{\delta}$ = 30.60 mas/yr, which is the
only extreme outlier in the two proper motion distributions: at 5.8$\sigma$ from the mean in RA 
proper motion and at 8.4$\sigma$ in DEC.

\begin{figure}[ht!]
\centering
\includegraphics[width=0.45\textwidth]{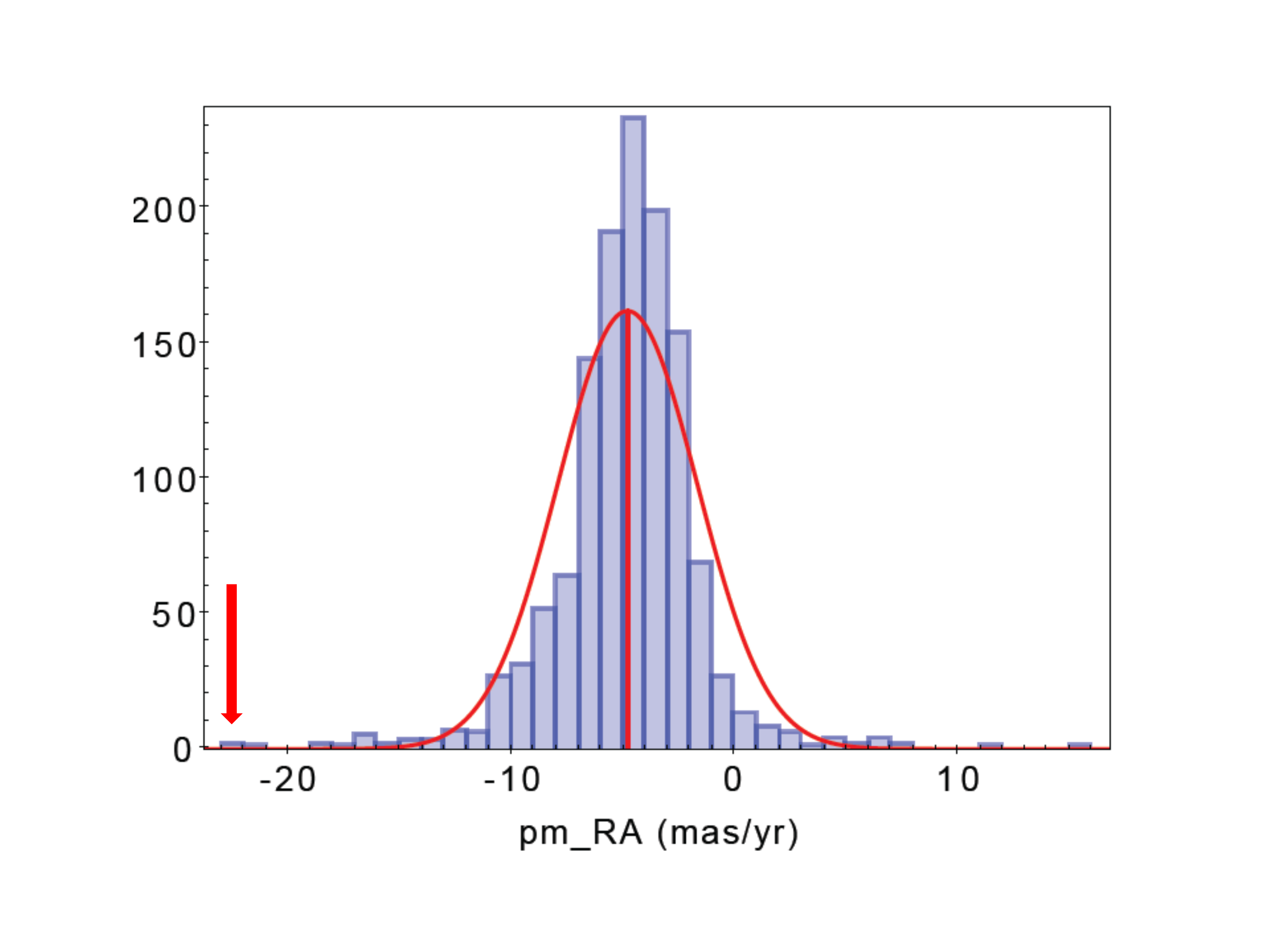}
\includegraphics[width=0.45\textwidth]{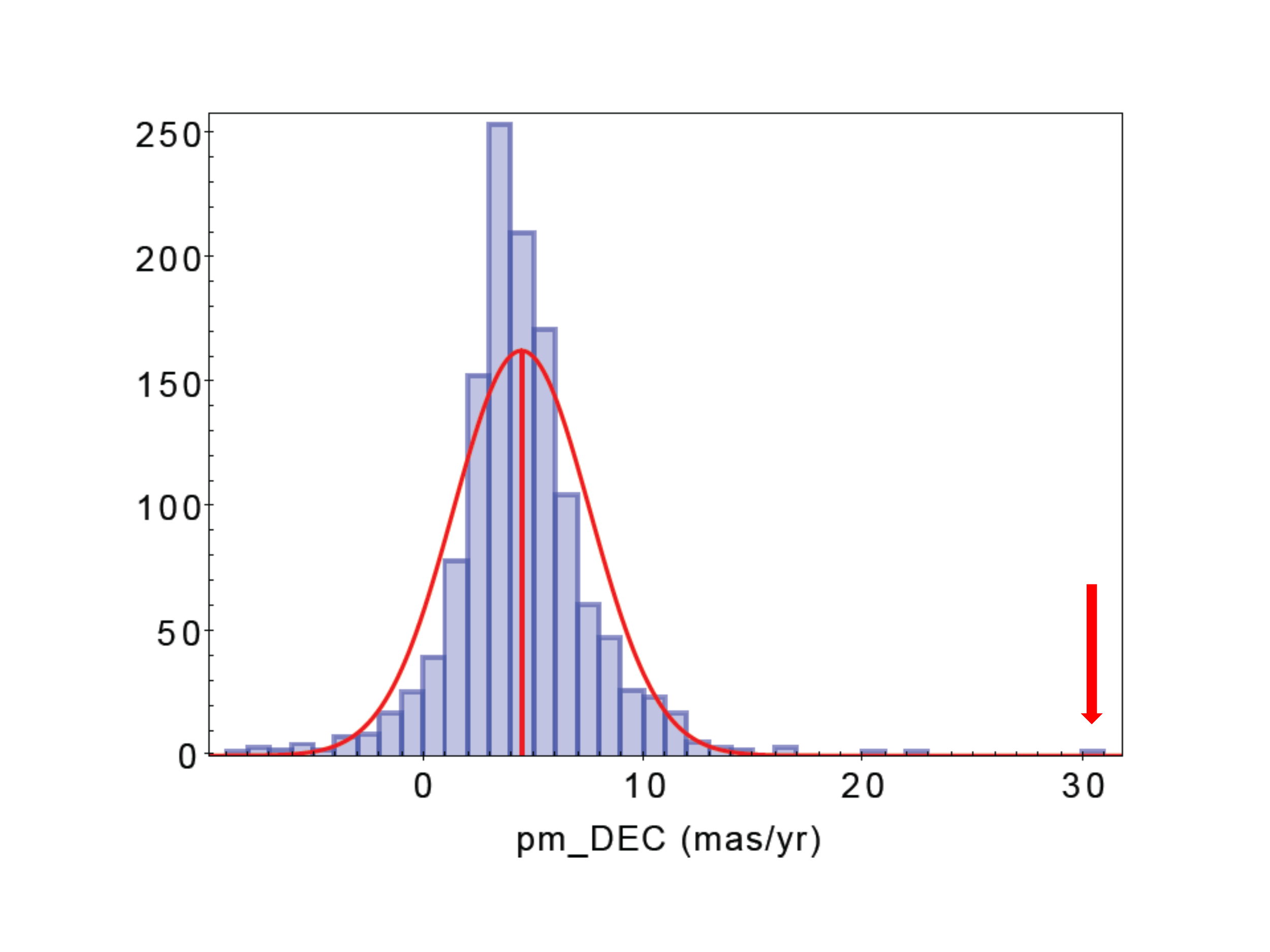}
\caption{Left: histogram of the distribution of proper motions in RA of
the stars in our sample. Right: same, for the proper motions in DEC. The 
Gaussian fits are overplotted in red.}
\label{Figure 3}
\end{figure}

\noindent
The distribution of total proper motions 
($\mu = \sqrt((\mu^{*}_{\alpha})^{2} + (\mu_{\delta})^{2})$) 
is shown in Figure 4 (left panel).
Since we know the distances to the stars, we can calculate their total 
velocities perpendicularly to the line of sight, $v_{tan}$ (using the 
expression: $v_{tan} = 4.7485\times\mu/\varpi$, $\varpi$ being the parallax). 
The resulting
distribution is shown in the right panel of Figure 4. 
Here we see again the same outlier, with a total proper motion of 38.15
mas/yr (that is 8.9$\sigma$ above the mean). The
distance to the star, from its parallax, is $d = 1.32^{+1.00}_{-0.39}$ kpc,
which gives a tangential velocity $v_{tan} = 239^{+181}_{-70}$ ~\kmso (5.4$\sigma$
above the mean).

\begin{figure}[ht!]
  \centering
\includegraphics[width=0.45\textwidth]{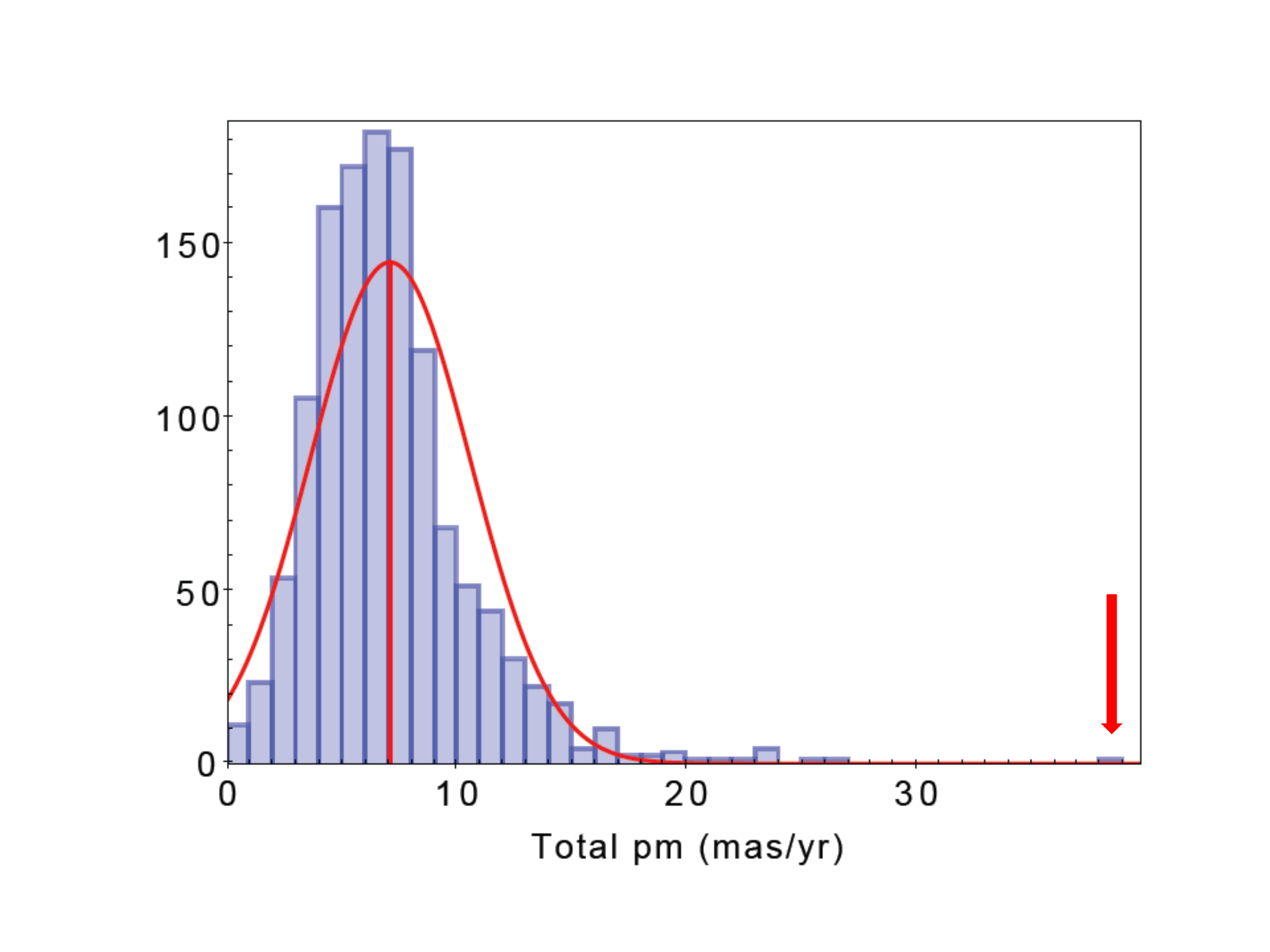}
\includegraphics[width=0.45\textwidth]{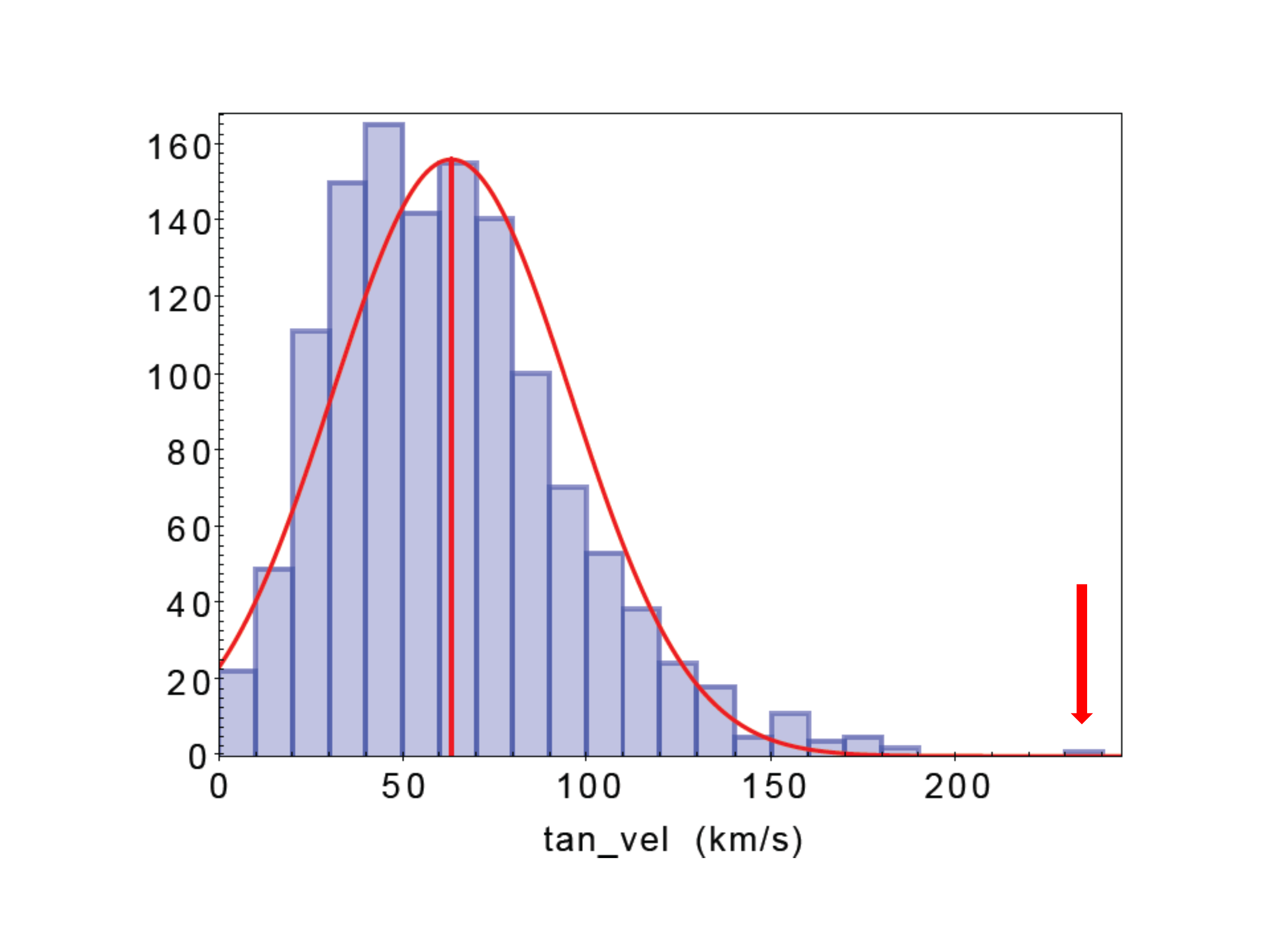}
\caption{Left: histogram of the distribution of the total proper motions of
  the sampled stars. Right: same for the velocities $v_{tan}$, 
perpendicular to the line of sight.}
\label{Figure 4}
\end{figure}

\noindent
Total speeds of that order are expected for MS or SG companions of a SN Ia
(Han 2008; see also Table 1 in Pan et al. 2014), so this star deserves 
further study. 

\noindent
Since the SNR G272.2-3.2 has a minimum age of $\sim$ 4,500 yr, any possible
surviving companion should have travelled an appreciable distance from the
site of the explosion, by now. From the proper motions measured, we can infer
its position at the time of the SN outburst. In Figure 5, that is made for
an age of 8,000 yr. We see that star  M\RomanNumeralCaps{5}-G272,
located at the periphery of the SNR at present, was very close to the center 
by then.   

\begin{figure}
\centering
\includegraphics[width=0.45\textwidth]{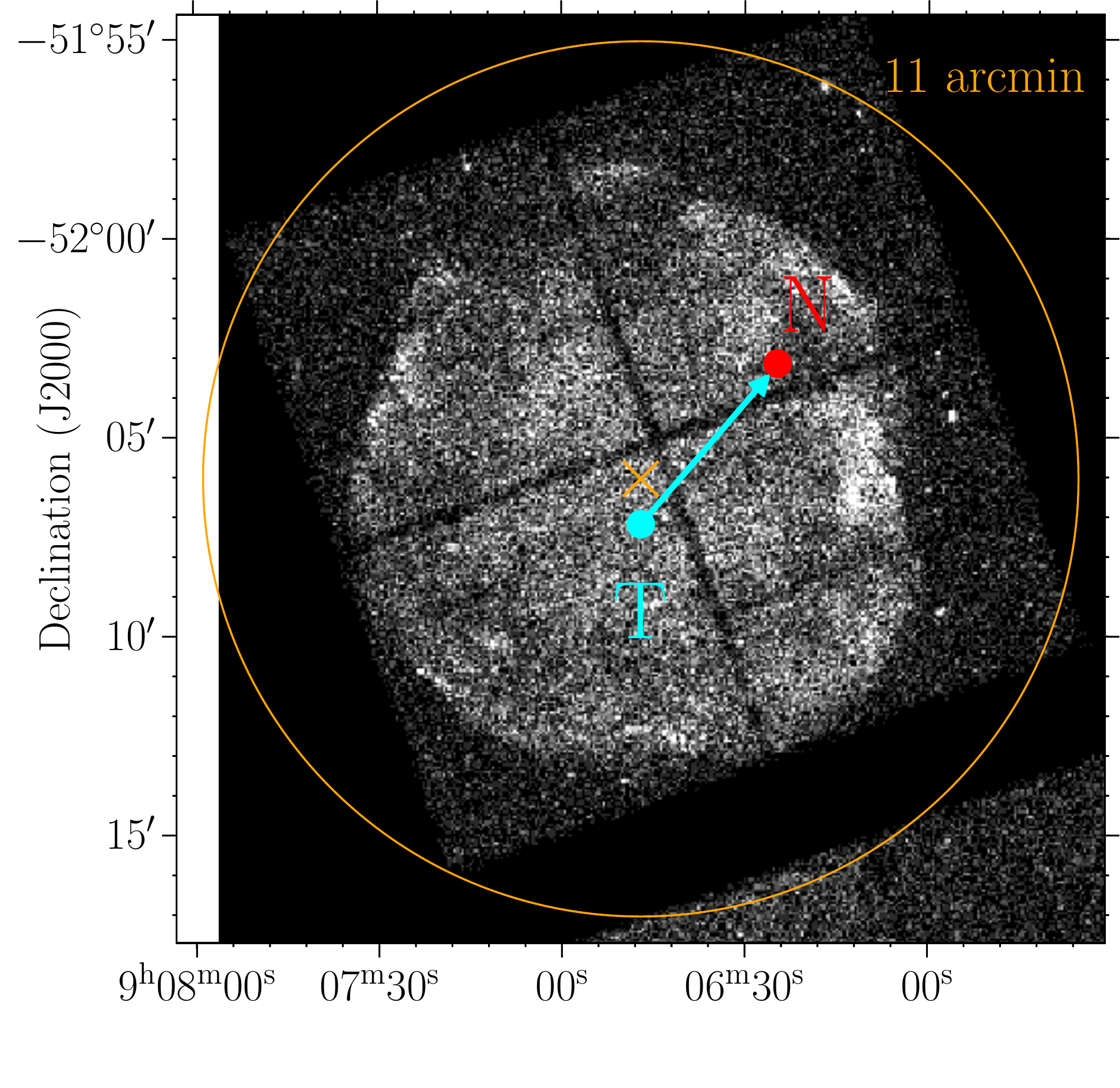}
\caption{Present position of the fastest moving star,
  M\RomanNumeralCaps{5}-G272 (red dot, labelled $N$), compared with
  that it had 8,000 yr ago (blue dot, labelled $T$). 
  The picture has been superposed to the Chandra image of the SNR in keV 
  range (S\'anchez-Ayaso et al. 2013). We see that the star, now nearing 
  the edge of the SNR, was close to the 
  center (marked with a red cross), then. The errors in the proper motions
  being very small, the uncertainty as to the past position is not larger
  than the size of the plotted point.
}
\label{Figure 5}
\end{figure}

\noindent 
The {\it Gaia} EDR3 does not provide information on the radial velocities
$v_{r}$ of any of those stars. From a spectrum of
M\RomanNumeralCaps{5}-G272 (see next
Section), we have measured  
a barycentric velocity $v_{bar}$ = 92.6$\pm$0.5 ~\kms ($v_{r}$ = 77.3 km ~\kmso 
in the LSR). That gives 
$v_{tot} = 256^{+181}_{-70}$ ~\kmso (barycentric), for this star.

\noindent
 It can be seen, from Figures 3 and 4, that there are a few stars at  
$\sigma >$ 3 in the corresponding distributions, but in view their positions 
relative to the SNR and their trajectories, they are not viable 
 candidates to be companions of the SN.

\subsection{The kinematics of star 
{\it Gaia} EDR3 5323900211541075328/MV-G272}

The star M\RomanNumeralCaps{5}-G272 looks like a
possible candidate to be
the surviving companion of the SN Ia that gave rise to SNR G272.2--3.2.
Subsequently, its kinematics is analyzed in more detail below. We will  
look at the
motion in Galactic coordinates taking for comparison the
current Besan\c con model of the Galaxy (https://model.obs-besancon.fr). Since 
the star
is a M1-M2 dwarf with solar metallicity (see next Section), we will only use,
from the model, M dwarfs whith the same metallicity and located at
distances 1 kpc $\leq d \leq$ 2 kpc, like the candidate star. The resulting 
distributions in $\mu^{*}_{l}$, $\mu_{b}$ and $v_{r}$ of the model stars 
are shown in Figure 6.

\noindent
With $\mu^{*}_{l}$ = -37.96 mas/yr, star M\RomanNumeralCaps{5}-G272 is 
at 6.1$\sigma$ above the mean, while its $\mu_{b}$ = 3.85 mas/yr is at
1.4$\sigma$ only, and $v_{r}$ = 77.1 ~\kmso in the LSR is at 2.4$\sigma$.
Thus, only the motion along the Galactic plane is really peculiar as compared 
with the Besan\c con model.

\begin{figure}[ht!]
  \centering
  \includegraphics[width=0.45\textwidth]{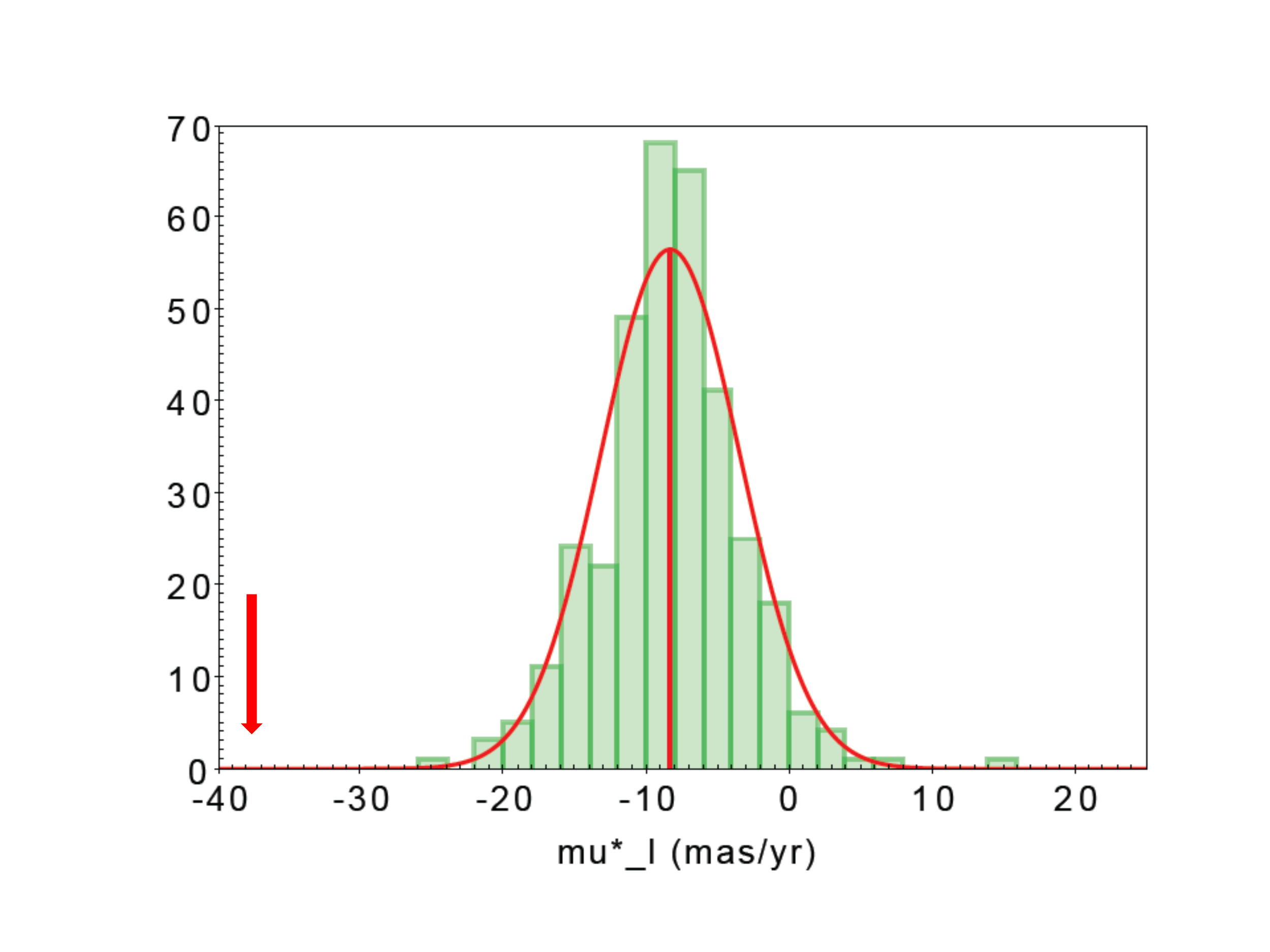}
  \includegraphics[width=0.45\textwidth]{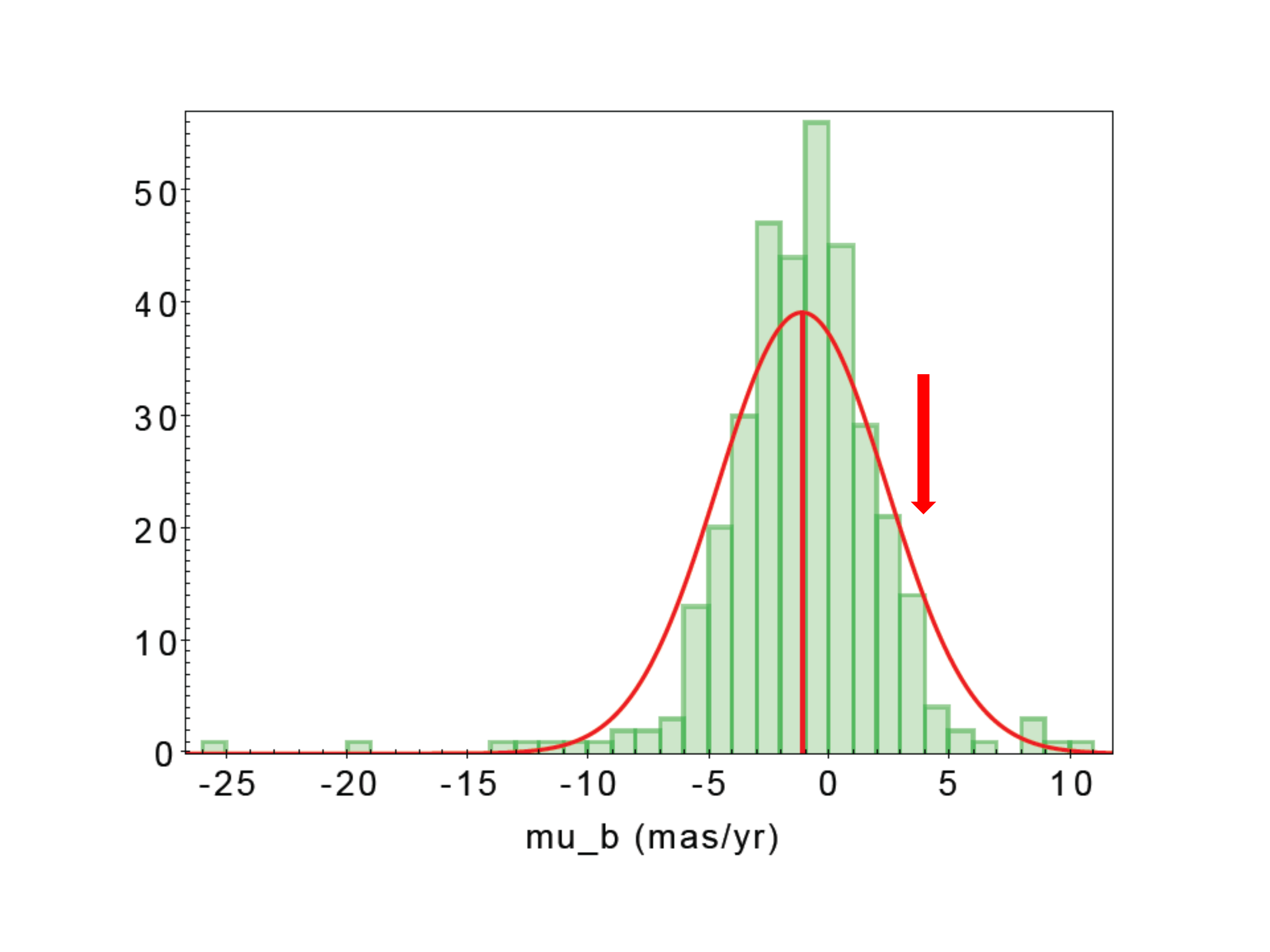}
  \includegraphics[width=0.45\textwidth]{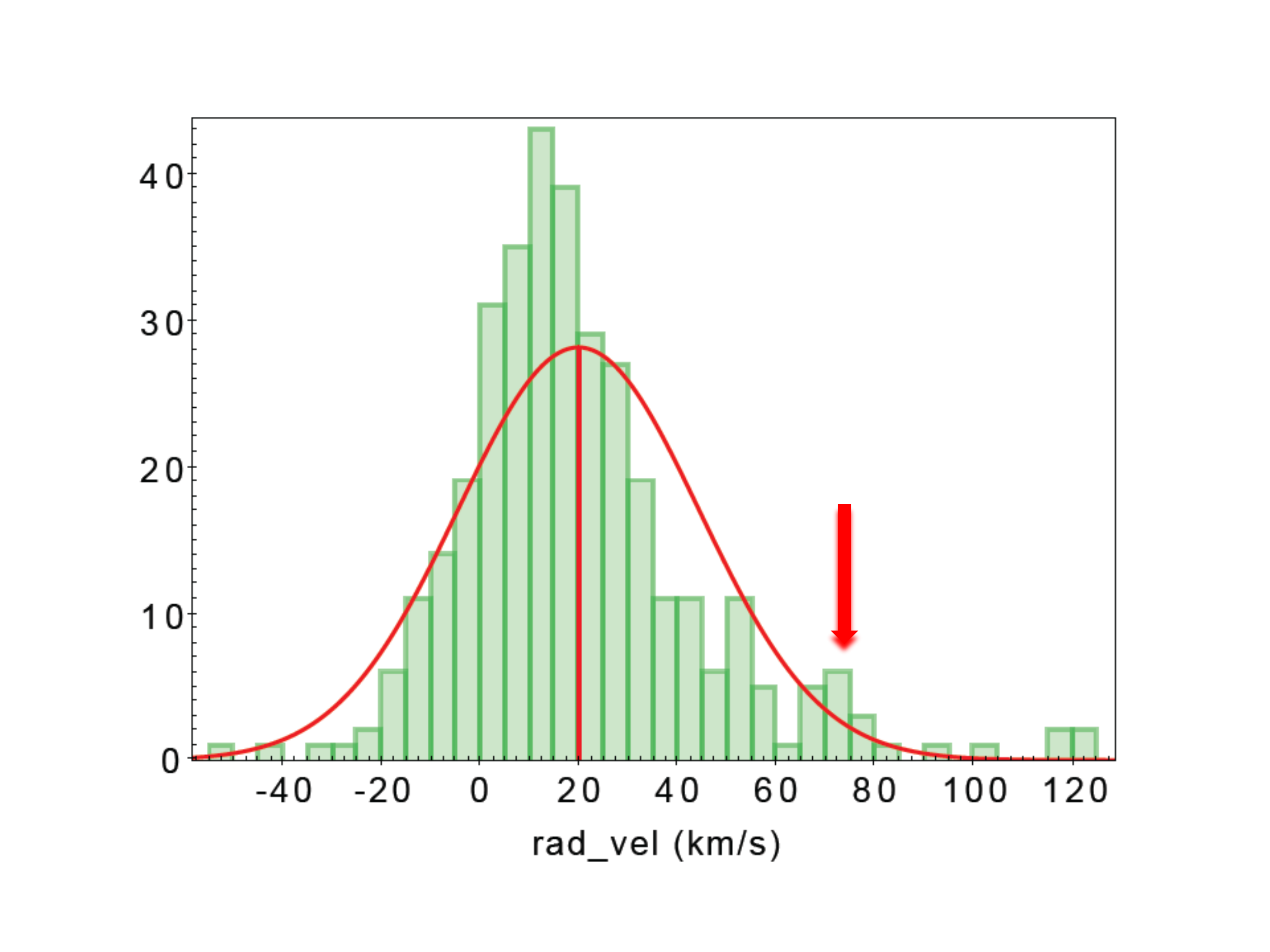}
  \caption{The distributions in $\mu^{*}_{l}$, $\mu_{b}$ (two upper panels) and
    $v_{r}$ (lower panel) of the
    M dwarf stars with about solar metallicity and at distances 1 kpc $\leq
    d \leq$ 2 kpc, in the direction of G272.2--3.2, from the Besan\c con model
    of the Galaxy (https://model.obs-besancon.fr; see text for details).}
  \label{Figure 6}
\end{figure}

\noindent
We can also select the M dwarf population within our sample. equally at
distances 1 kpc $\leq d \leq$ 2 kpc, and look at its
distribution in $\mu_{l}$ and $\mu_{b}$. In this way we know how star
M\RomanNumeralCaps{5}-G272 moves as compared with the surrounding
stars of the
same type. The distributions are shown in Figure 7. This star is a clear
outlier in $\mu^{*}_{l}$, at 7.4$\sigma$ above the mean, while it is only at
1.7$\sigma$ in $\mu_{b}$,

\begin{figure}[ht!]
  \centering
  \includegraphics[width=0.45\textwidth]{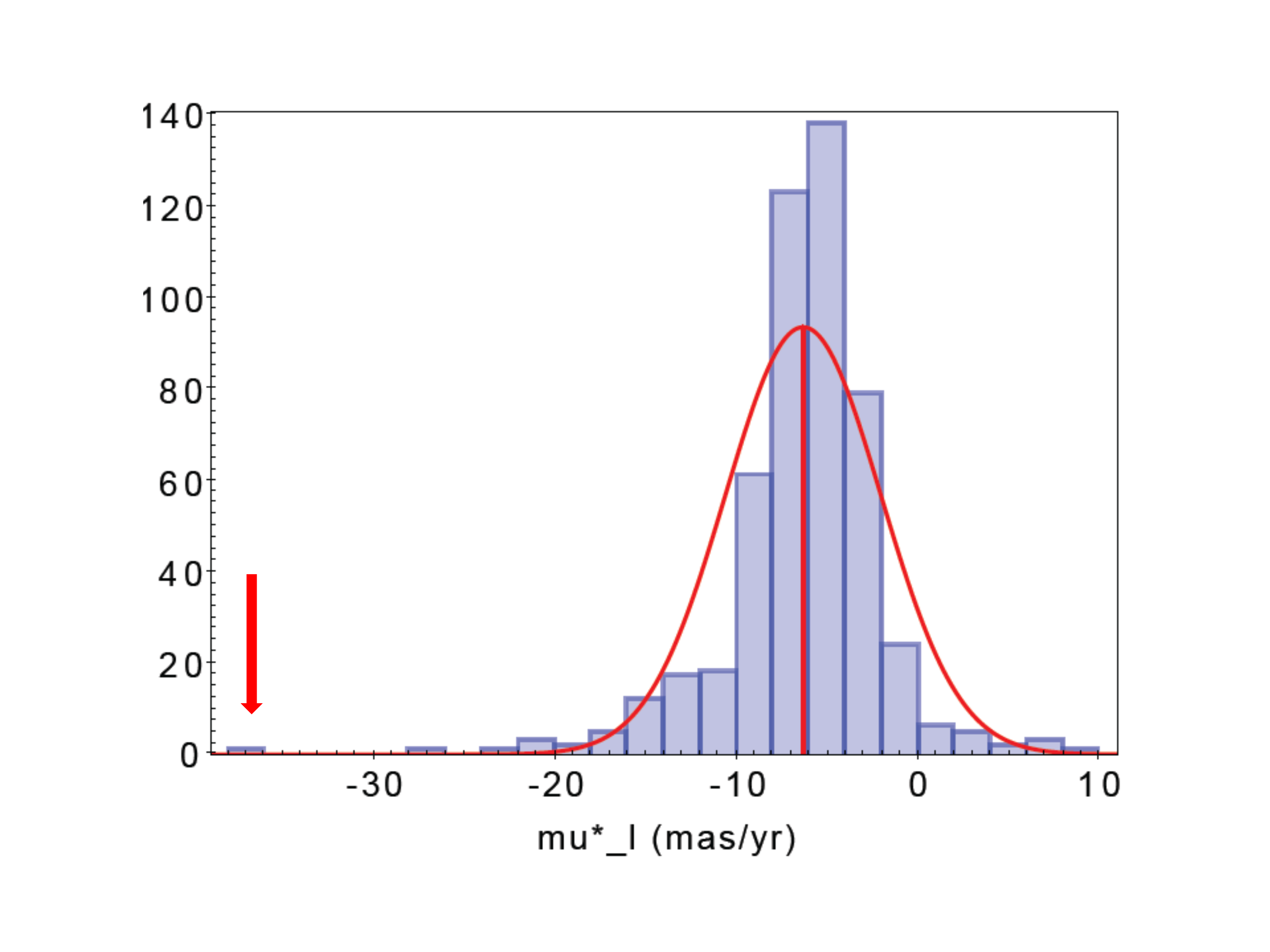}
   \includegraphics[width=0.45\textwidth]{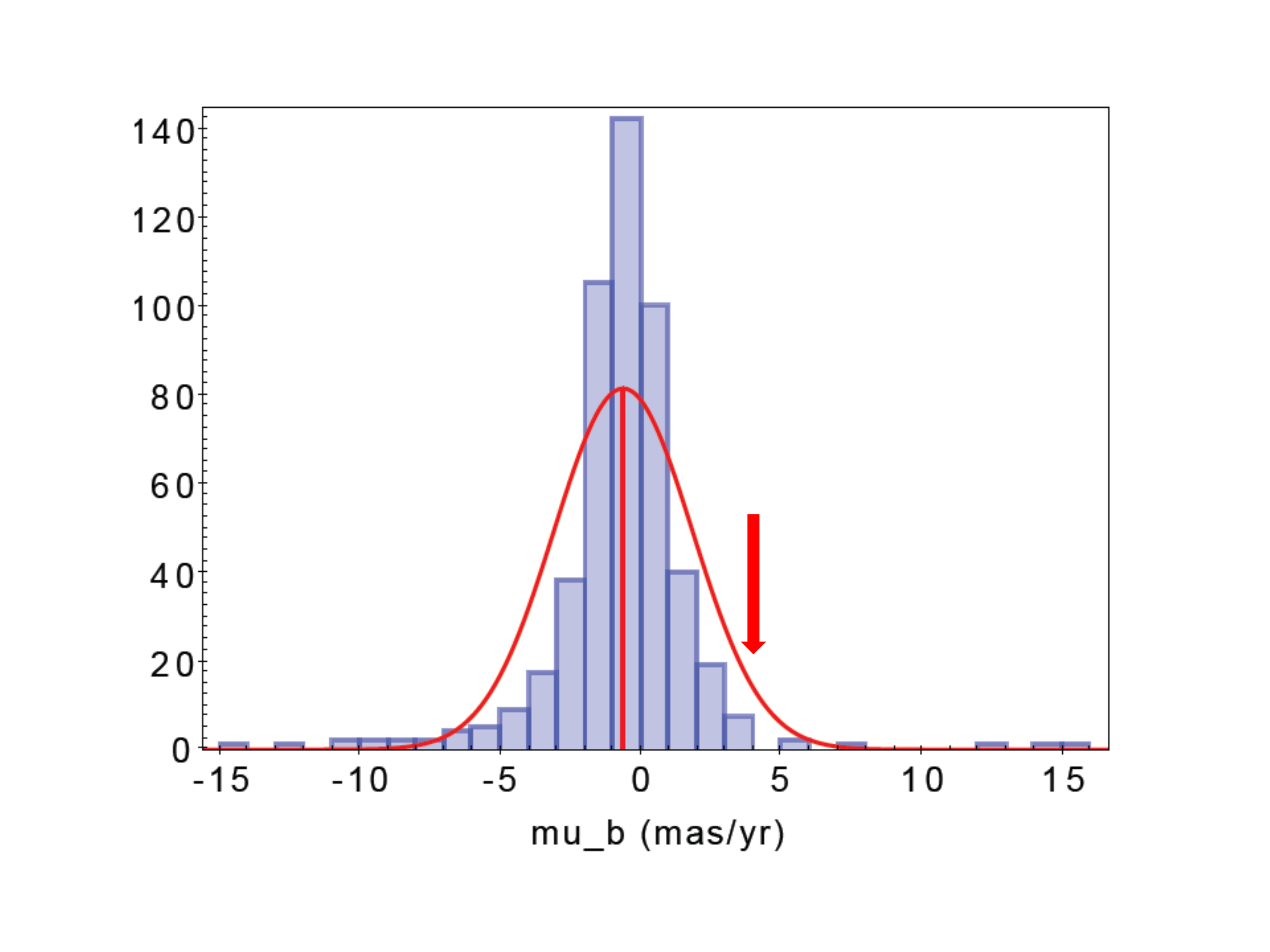} 
   \caption{The distributions in $\mu^{*}_{l}$ (left panel) and $\mu_{b}$ (right
     panel) of the M dwarf stars at distances 1 kpc $\leq d \leq$ 2 kpc in our
     sample  from {\it Gaia} EDR3.}  
  \label{Figure 7}
  \end{figure}

\noindent
To investigate further the kinematics of star
MV-G272, we have also
calculated its
orbit, using {\it GravPot16} (Fern\'andez--Trincado 2017). In Figure 8 we
show the 3D orbit and its projections on the XY, XZ and YZ planes of the
referential system of the Galaxy. We see that the projection of the orbit on
the XY plane is very eccentrical, with $e$ = 0.447. Only 13 stars in our
sample have radial velocities determined in {\it Gaia} EDR3, and only them 
can thus be used to calculate eccentricities. The mean
eccentricity is $\langle e \rangle$ = 0.167, with $\sigma$ = 0.063. Therefore,
that of star M\RomanNumeralCaps{5}-G272 is 4.5$\sigma$ above the mean,
as illustrated in Figure 9.

\begin{figure}[ht!]
  \centering
  \includegraphics[width=0.35\textwidth]{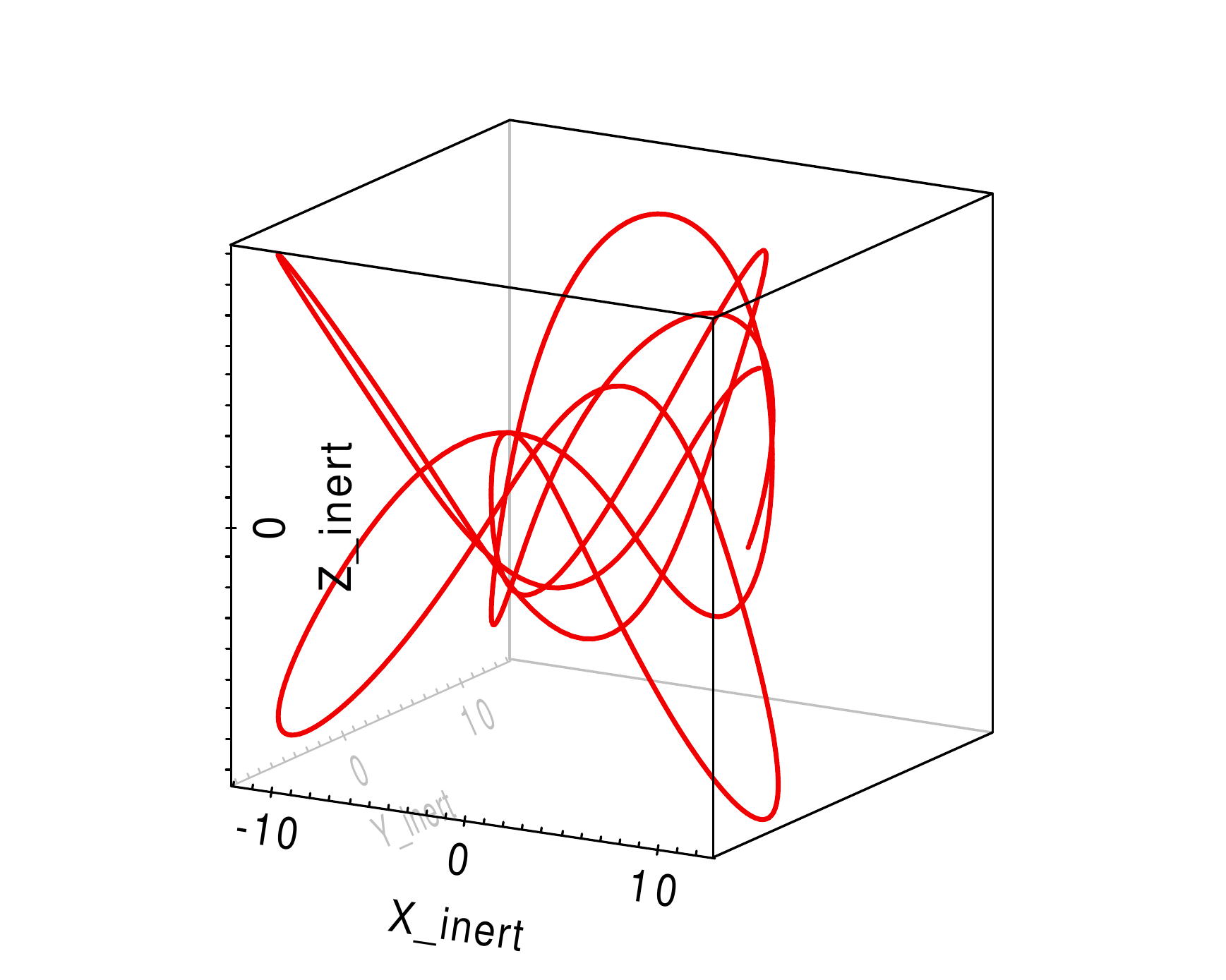}
  \includegraphics[width=0.3\textwidth]{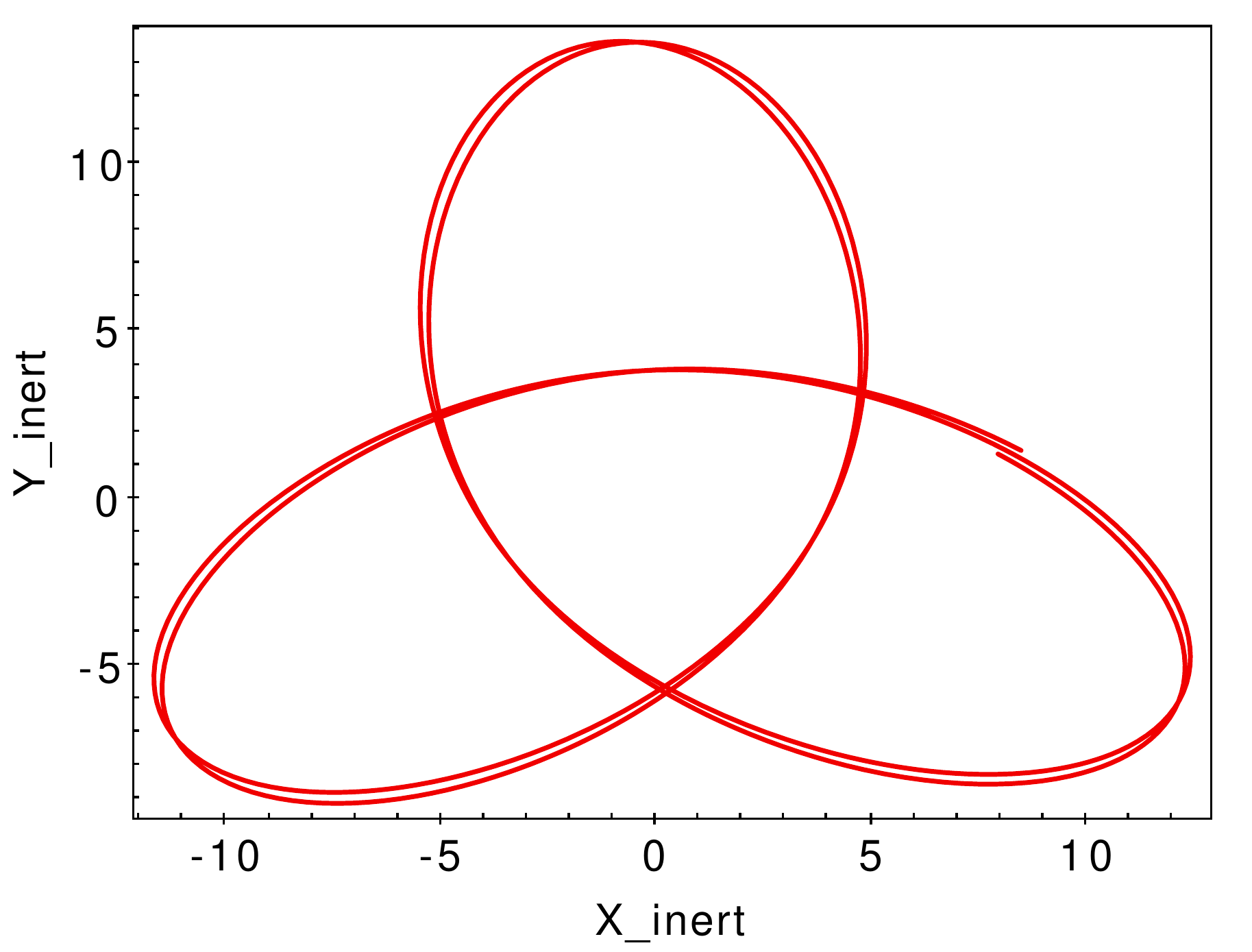}
  \includegraphics[width=0.3\textwidth]{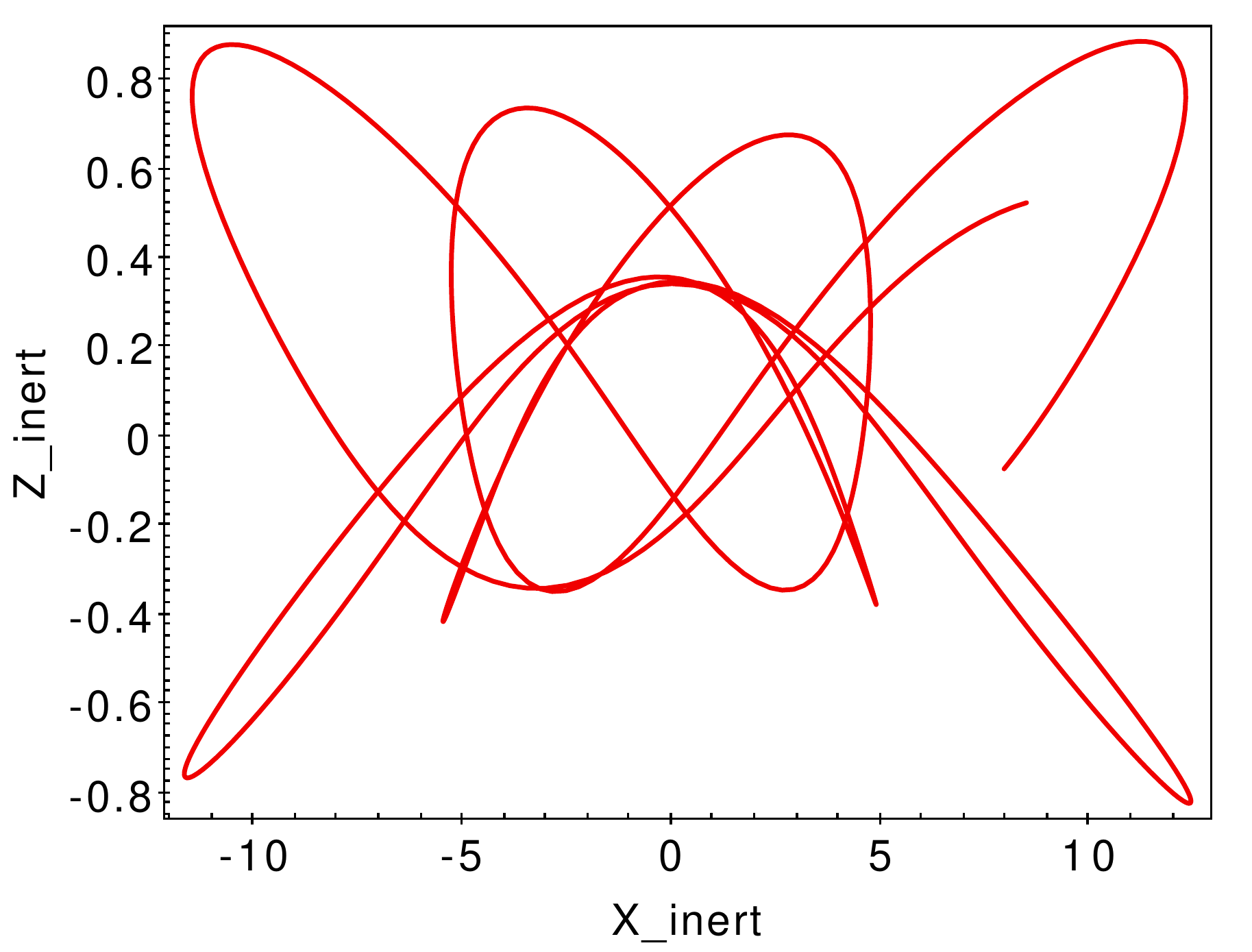}
    \includegraphics[width=0.3\textwidth]{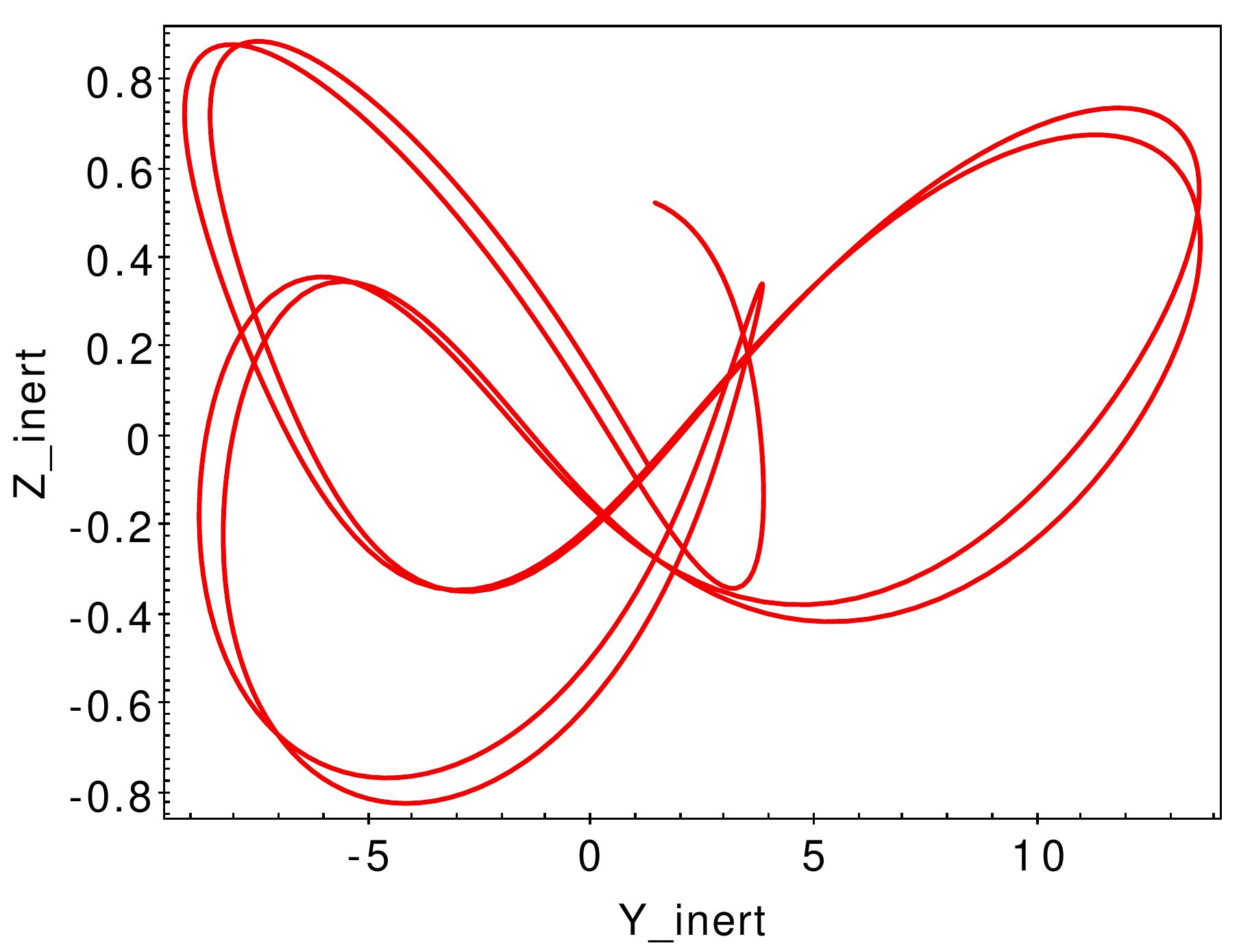}
    \caption{3D orbit of star M\RomanNumeralCaps{5}-G272 along
      10$^{9}$ yr (forward in time) and its 
      projections on the XY, XZ and YZ planes of the referential system
      of the Galaxy.}
    \label{Figure 8}
\end{figure}

\begin{figure}[ht!]
  \centering
  \includegraphics[width=0.3\textwidth]{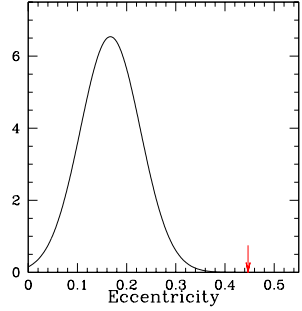}
  \caption{The eccentricity distribution of the orbits of the 13 stars in
    our sample having their radial velocities measured in {\it Gaia} EDR3.
  An arrow marks that of star M\RomanNumeralCaps{5}-G272.}
  \label{Figure 9}
  \end{figure}

\section{Reddening and spectroscopy}

\begin{figure}[ht!]
   \centering
   \includegraphics[width=0.60\textwidth]{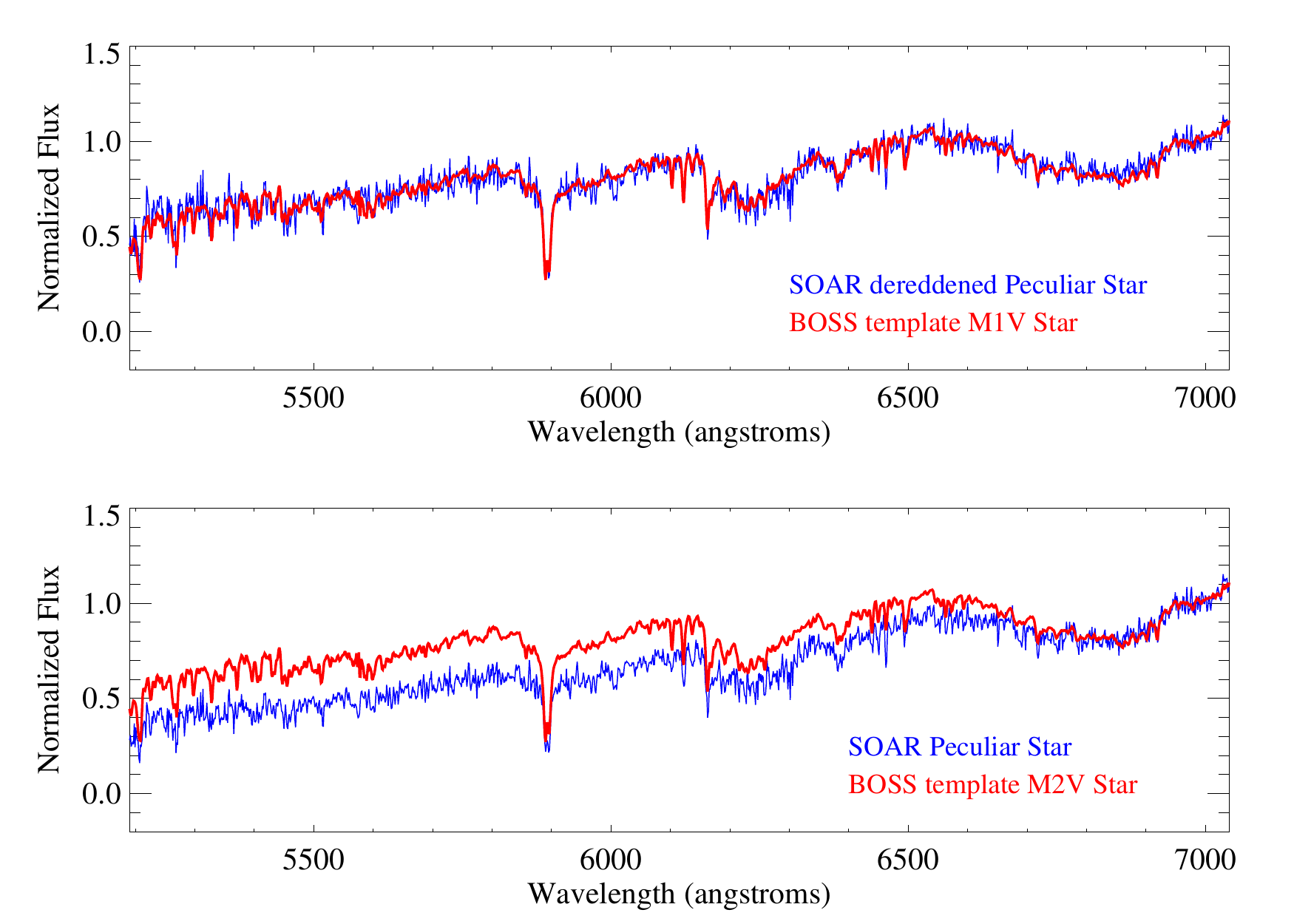}
      \includegraphics[width=0.60\textwidth]{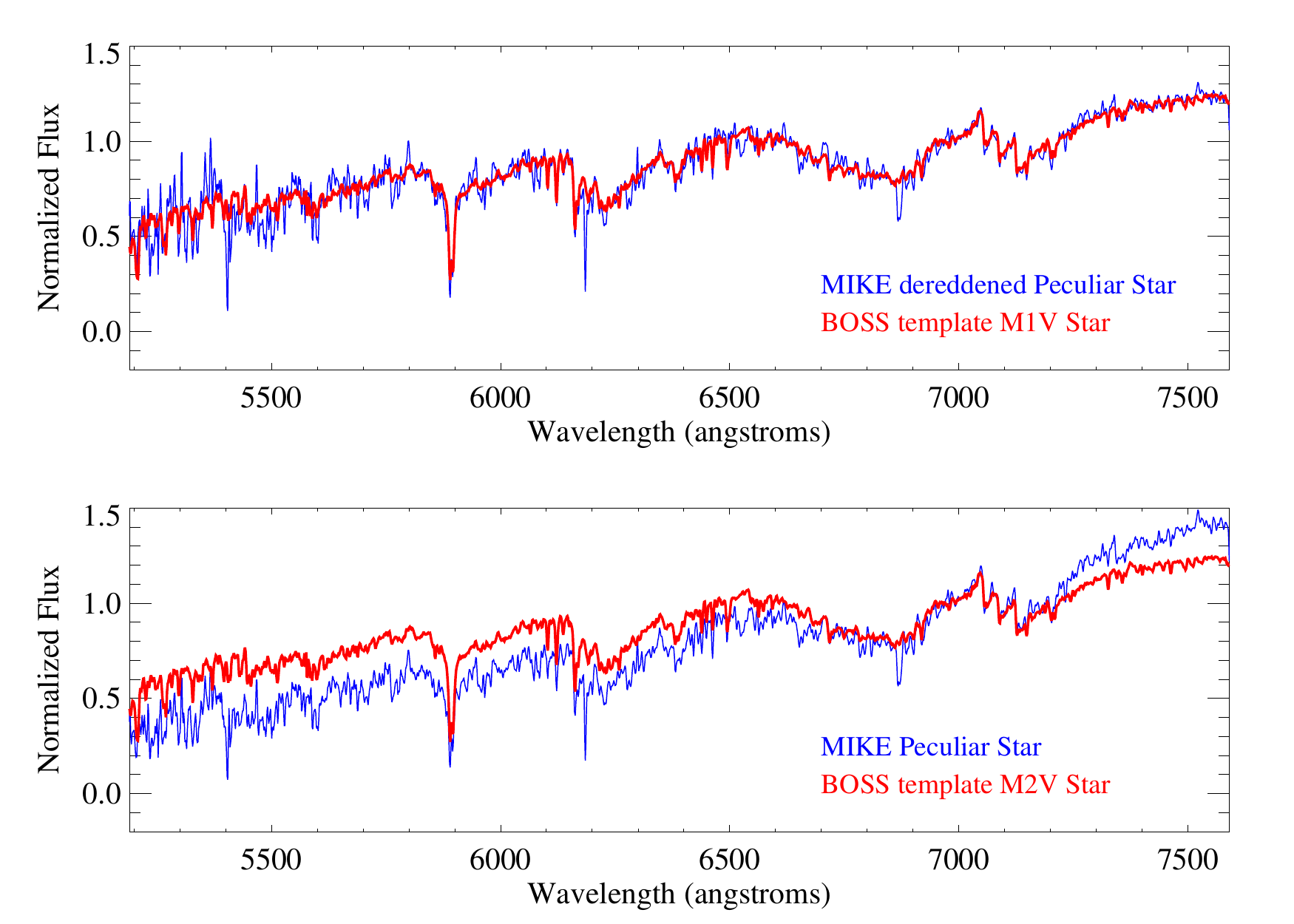}
      \caption{Two upper panels: spectrum of star
        {\it Gaia} EDR3 5323900211541075328 (M\RomanNumeralCaps{5}-G272), taken 
      with the 
      {\it Goodman} spectrograph at the 4.1m {\it SOAR} telescope. In the 
      first panel, it is corrected for a reddening {\it E(B - V)} = 0.532 and 
      superposed to the template for a M1 dwarf with solar metallicity, 
      while in the second one it is compared with the template for 
      a M2 dwarf with the same metallicity but without correction for 
      reddening there.
      Two lower panels: spectrum of the same star, taken with the {\it MIKE}
      spectrograph at the 6.1m {\it Clay} telescope, covering
      a similar wavelength range. In the first panel, it is  superposed to the 
      template for a M1V 
      star with solar metallicity and corrected for a reddening 
      {\it E(B - V)} = 0.532. In the second panel, it is superposed to the 
      template for a M2V star, also with solar metallicity, without
      correcting for reddening.
      Observed and template spectra have been normalized by dividing by
      a constant equal to the mean value of the fluxes in the spectral range 
      6,950-7,000 \AA. The {\it MIKE} spectrum has been degraded to a resolution
      R$\sim$2,000, analogous to that of the {\it BOSS} templates used by the 
      {\it PyHammer} code.
      All spectra have been sampled with a pixel size of 1.298 \AA \ per pixel.
      }
    \label{Figure 10}
  \end{figure}

\noindent
Measured
{\it Gaia} magnitudes are affected by absorption due to 
intervening gas and dust. In order to compare the observed photometry with
theoretical models of the evolution of possible stellar companions after the
SN Ia explosion, the extinction in different bands must be accurately
estimated and substracted. 

\noindent
There are only four stars in our sample which have their parameters 
$T_{eff}$, log $g$ and [Fe/H] determined in the {\it Gaia} EDR2 (see Table 2).

   \begin{table*}
      \centering
      \caption{
        {\it Gaia} EDR2 stars used for the reddening estimate}
      \begin{tabular}{lcccccc}
   \tableline  
    \tableline
{\it Gaia} ID & $G$ & $G_{BP}$ & $G_{RP}$ & T$_{\rm eff}$ & log $g$ & [Fe/H] \\
\tableline
5323849427851815808 & 12.26 & 13.31 & 11.25 & 5000 & 3 & 0 \\
5323897355382648960 & 12.65 & 13.43 & 11.77 & 5000 & 3 & 0 \\
5323899352554079101 & 13.26 & 14.30 & 12.26 & 4500 & 3 & 0 \\
5323899352554079104 & 13.25 & 14.28 & 12.25 & 4500 & 3 & 0 \\
\tableline
\end{tabular}
\end{table*}

Using the
expressions by Carrasco\footnote{{\it Gaia Data Release Documentation} 5.3.7; 
see also Jordi et al. (2010)}, we translate their magnitudes to the Johnson's 
UBVRI system and then
compare the measured colors with those corresponding to stars with the same
parameters (Hundaselt et al. 2000). From the color excesses,
we deduce the extinctions $A_{V}$, $A_{R}$, $A_{I}$ for each of the four stars.
Since the stellar parameters from {\it Gaia} EDR2 do not seem 
very accurate, we
do not expect a perfect coincidence in the values obtained for the different
stars, but the average value for $A_{V}$ is 1.65 mag, in agreement with 
that obtained from the spectral fits below.

\noindent
Another estimate comes from fitting spectra of star 
M\RomanNumeralCaps{5}-G272, taken with the $MIKE$ spectrograph at
the 6.5m $Clay$ telescope and with the {\it Goodman} spectrograph at the 4.1m
{\it SOAR} telescope (both covering similar wavelength ranges).
We have used the {\it PyHammer} tool (Kesseli et al. 2017, 2020) to infer that 
the best fit spectral type is M1V-M2V. This code uses a set of templates for 
different spectral types and luminosity classes with a discrete set of 
metallicity values with a step of $\sim$0.5 dex, created from observed 
SDSS/BOSS spectra at R$\sim$2,000.  In Figure 10  are shown the fits of
both spectra to the template for a M1V
star, with solar metallicity and a reddening of {\it E(B-V)} = 0.532 (two
top panels), and to a M2V star, with negligible reddening and
solar metallicity too (bottom panels). Good fits to a M1V spectrum can also
be obtained for smaller redddening and a somewhat higher metallicity ([Fe/H] =
+0.5). We prefer the first fits, which correspond, for $A_{V}$ = 3.1$\times$
E(B -V), to $A_{V}$ = 1.65 mag, and coincides with the 
estimate made from photometry.

\noindent
As mentioned above, we have measured a radial velocity of
$v_{r}$ = 92.6$\pm$0.5 ~\kmso 
(barycentric) or $v_{r}$ = 77.3$\pm$0.5 ~\kms (LSR),
using the {\it MIKE} spectrum.

\noindent    
A deeper and more complete analysis of the {\it MIKE} spectrum is made in 
the next Section.

\begin{figure}[ht!]
\centering
\includegraphics[width=0.95\textwidth,angle=0]{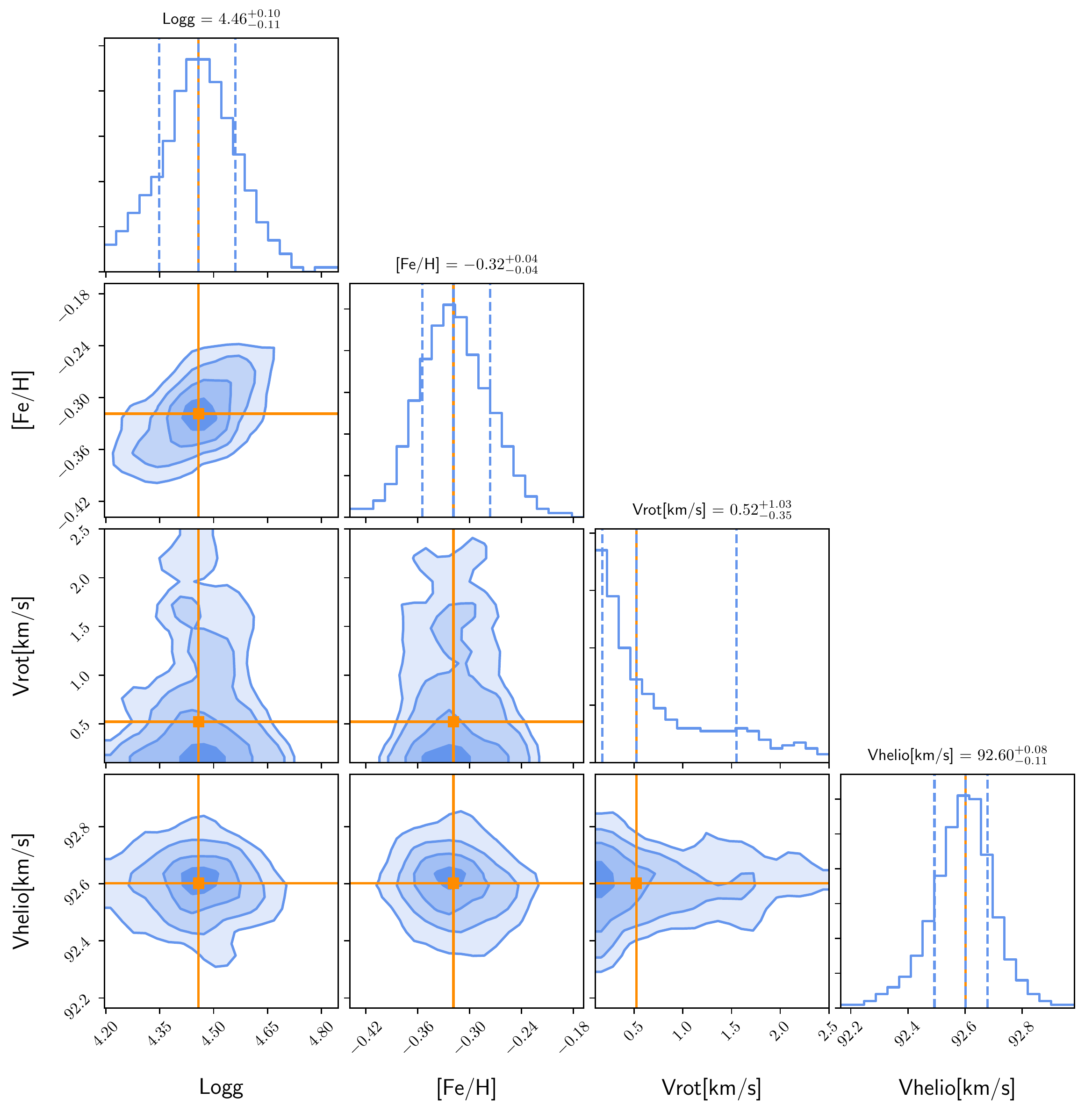}
\bigskip
\noindent
\begin{minipage}[b]{0.65\textwidth}
\caption{ 
Posterior distributions of the parameters of the analysis of the {\it MIKE} 
spectrum 
using our Bayesian code to derive the effective temperature $T_{\rm eff}$, 
surface gravity $\log g$, metallicity [Fe/H], rotational velocity 
$V_{\rm rot}$, and  heliocentric radial velocity  $V_{\rm helio}$. A fixed 
$T_{\rm eff}$ of about 3800 K has been assumed here.}
\end{minipage}
\label{figdist0}
\end{figure}

\section{Stellar parameters and metallicity of {\it Gaia} EDR3 
5323900211541075328 (star MV-G272)}

We have analysed the high-resolution {\it MIKE} spectrum ($R\sim 28,000$) to 
try to 
estimate global metallicity and some element abundances, from individual lines 
available in the red part of the {\it MIKE} (Bernstein et al. 2003) spectrum. 
The star is quite faint, which gives 
an estimated  S/N$\sim 18$ at 7,500~{\AA}. The spectrum has a total 
exposure time of 13,200~s. For comparison we also analysed two high-resolution 
CARMENES VIS spectra (Reiners et al. 2018) of two stars classified as M1V 
stars and the solar 
ATLAS spectrum~ (Kurucz et al. 1984) as a reference (see Appendix and 
discusssion below).
The two CARMENES spectra of these stars have been recently analysed (Marfil et
al. 2021) with the
SteParSyn code (Tabernero et al. 2022, 2021), providing the following set of 
parameters:  
$T_{\rm eff}/\log g/$[Fe/H]~$ = 3,603 / 4.99 / -0.52$ for star Karmn J00183+440 
(GX And) and $T_{\rm eff}/\log g/$[Fe/H]~$ = 3,825 / 4.94 / -0.04$ for 
star Karmn J05415+534 (HD 233153). Deriving the metallicities of M dwarfs even 
with high quality spectra at high resolution is a challeging exercise, with 
differences of 0.3~dex from different methods (Passeger et al. 2022). 
These analyses have been 
done from individually resolved lines at very high resolution of these very 
high quality spectra, using those codes mentioned above.

\noindent
We have also analysed single CARMENES VIS spectra of these two stars and the 
solar ATLAS 
spectrum, all degraded to a resolving power of 28,000 and with injected noise 
to  S/N~$\sim 18$ at 7,500~{\AA}, to match the {\it MIKE} spectrum 
resolution and quality. 
Thus, we implemented a Bayesian Python code that compares the observed spectrum 
with a synthetic spectrum in the spectral range 7,000-8,750 \AA.  
Both the observed spectra and the synthetic spectra are normalized using a 
running mean
filter with a width of 200 pixels at a dispersion of 0.069{\AA} per pixel
(see Figures in Appendix).
We performed a Markov Chain Monte Carlo (MCMC) with 5,000 chains implemented 
in {\sc emcee} (see Foreman-Mackey et al. 2013), sufficient to get a 
statistically significant result.
We use a small 3x3x3 grid of synthetic spectra with values 
$T_{\rm eff}/\log g/$[Fe/H] of $3500 - 4500 / 3.0 - 5.0 / -1.0 - 0.5$ and steps 
of 250~K / 0.5~dex / 0.5~dex, computed with the
{\sc SYNPLE}\footnote{Available at https://github.com/callendeprieto/synple} 
code, assuming a microturbulence $\xi_{\rm mic}=0.8$~\kmso, and ATLAS9 model 
atmospheres with solar $\alpha$-element abundances ([$\alpha$/Fe]=0, see 
Castelli \& Kurucz 2003). Details of the fits are shown in the Appendix.

\noindent
The model includes as free parameters, the effective temperature, $T_{\rm eff}$,
surface gravity, $\log g$, [Fe/H], rotational velocity, 
$V_{\rm rot}$, and relative radial velocity, $V_{\rm rel}$ (Figure 11).
We run a simulation 
leaving free all the parameters and found that the simulation converges to a 
$T_{\rm eff}$ value at the lower edge of the grid. A similar result with a 
lower $T_{\rm eff}$ value by about 250K than those obtained at high-resolution 
was obtained when analysing the two CARMENES spectra.
Thus we decided to run a simulation by fixing the $T_{\rm eff}$ at 3800~K for 
our target star. The posterior distributions of the simulation is displayed
in Figure 11, that provide the values 
$\log g/$[Fe/H]~$ =  4.46 / -0.32$. For the two degraded 
CARMENES spectra we got $T_{\rm eff}/\log g/$[Fe/H]~$ = 3,580 / 4.37 / -0.69$ 
for star Karmn J00183+440 (GX And) and 
$T_{\rm eff}/\log g/$[Fe/H]~$ = 3,805 / 4.67 / -0.04$ for 
star Karmn J05415+534 (HD 233153). As seen in Figure 12, the 
rotational velocity is not resolved with the instrumental full width at half 
maximum FWHM of 10.7~\kmso, providing a value consistent with zero and an 
upper-limit at 3$\sigma$ of $V_{\rm rot}< 3$~\kmso. The heliocentric radial 
velocity of the star is estimated at 92.60  $\pm$ 0.5  ~\kms, which
translates into 
$V_{\rm LSR} = 77.3 \pm 0.5$~\kms in the LSR. 

\noindent
We analysed the solar ATLAS spectra to get the reference solar element 
abundances, including all the relatively isolated features available in the 
{\it  MIKE} spectrum of our target star, including five Fe lines, seven Ti 
lines, two 
Cr and Na lines and one Ni and Al lines. For that exercise we used a grid of 
models with values $T_{\rm eff}/\log g/$[Fe/H] of 
$5,500 - 6,000 / 4.0 - 5.0 / -0.5 - 0.5$ and steps of 
250~K / 0.5~dex / 0.5~dex, assuming a fixed $\xi_{\rm mic} = 0.95$~\kmso. 
The analysis of the degraded solar spectrum provided all the element 
abundances within [X/H]~$= -0.13$ for Ti and $-0.03$~dex for Na. The Ca 
features were discarded because they provided very different results for Ca~I 
and Ca~II lines even in the solar case.
The tentative abundance ratios found in the target star are [Na/H]~$=-0.10$, 
[Al/H]~$=-0.23$, [Ti/H]~$=-0.05$, [Cr/H]~$=-0.08$, and [Ni/H]~$=-0.23$. 
These values may indicate a slight enhancement at 0.1--0.2 dex in all 
the element abundances with respect to the metallicity value of 
[Fe/H]~$=-0.32$. We consider them very tentative with this methodology in M 
dwarfs given also the S/N of the {\it MIKE} spectrum. In any case, 
these stars being almost fully convective, any captured material from the
SN ejecta should be strongly diluted. 

\noindent
    The charateristics of star M\RomanNumeralCaps{5}-G272 are
    summarized in Table 3.

\begin{figure}[ht!]
\centering
\includegraphics[width=0.95\textwidth,angle=0]{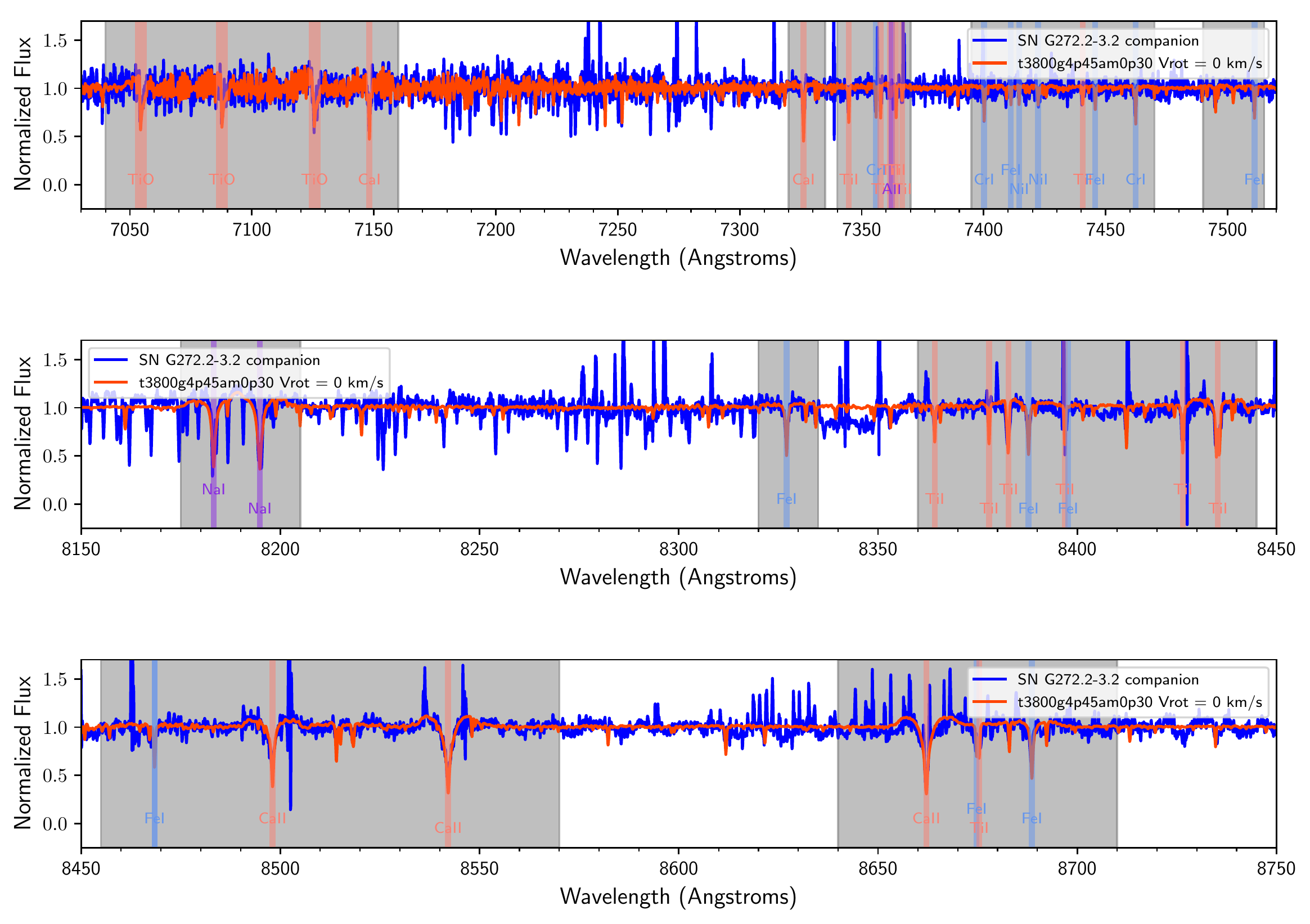}
\bigskip
\noindent
\begin{minipage}[b]{0.95\textwidth}
\caption{Normalized {\it MIKE} 1D spectrum of star Gaia EDR3 
5323900211541075328 (M\RomanNumeralCaps{5}-G272), corrected for 
barycentric radial velocity, and normalized to unity using a running mean 
filter with a width of 200 pixels at 0.069~{\AA} per pixel, with a 
signal-to-noise ratio of  $\sim$~18 at 7,500~{\AA}.
We also display an interpolated SYNPLE synthetic spectrum with the stellar 
parameters $T_{\rm eff}=3,800$~K, $\log g = 4.45$ and and metallicity 
[Fe/H]~$= -0.3$. The regions used to estimate the metallicity are shown in 
grey and the different lines used for chemical analysis are also highlighted.}
\end{minipage}
\label{figspec1}
\end{figure}

    \begin{table}[h!] 
      \centering
      \caption{Characteristics of star
        {\it Gaia} EDR3 5323900211541075328 (M\RomanNumeralCaps{5}-G272)}
      \begin{tabular}{lcc}
   \tableline  
    \tableline
 
    $\mu^{*}_{\alpha}$ (mas/yr) & $\dots$ & -22.79 \\
    $\mu_{\delta}$ (mas/yr)    & $\dots$ & 30.60 \\
    $\mu$ (mas/yr)           & $\dots$ & 38.15 \\
    $\mu^{*}_{l}$ (mas/yr)     & $\dots$ & -37.96 \\  
    $\mu_{b}$ (mas/yr)        & $\dots$ &  3.85 \\
    $d$ (kpc)                & $\dots$ & 1.32$^{+1.00}_{-0.39}$   \\
    $v_{\rm tan}$ (~\kmso)       & $\dots$ & 239$^{+181}_{-70}$ \\
    $v_{r}$ (~\kmso)           & $\dots$ & 77.3$\pm$ 0.5 (LSR)  \\
    $v_{r}$ (~\kmso]           & $\dots$ & 92.6$\pm$ 0.5 (barycentric) \\
    $v_{\rm tot}$ (~\kmso)      & $\dots$ & 256$^{+181}_{-70}$ \\
    $G$ mag                 & $\dots$ & 19.85  \\
    $G_{BP}$ mag             & $\dots$ & 21.03  \\
    $G_{RP}$ mag             & $\dots$ & 18.77  \\
    Spectral type           & $\dots$ & M1-M2  \\
    Luminosity class        & $\dots$ & V      \\
    $[Fe/H]$                 & $\dots$ & -0.32$\pm$0.04 \\
    $M$ ($M_{\odot}$)         & $\dots$ & 0.44-0.50    \\
    $R$ ($R_{\odot}$)         & $\dots$ & 0.446-0.482   \\
    $T_{\rm eff}$ (K)         & $\dots$ & 3,600-3,850 \\
    log $g$                  & $\dots$ & 4.46$^{+0.10}_{-0.11}$ \\
    log($L/L_{\odot}$)        & $\dots$ & -1.54/-1.39  \\

    \tableline
      \end{tabular}
      \end{table}

\section{Star {\it Gaia} EDR3 5323900211541075328/MV-G272 
as a possible SNIa companion}

We know that, at present, star  M\RomanNumeralCaps{5}-G272 has the
characteristics of a M1-M2 dwarf. Assuming that it were the companion of the 
SNIa that gave rise to the SNR G272.2-3.2 and that its mass  and radius 
was similar to the present ones, at the time of the explosion, then it should  
have been in close orbit with a 1.4 $M_{\odot}$ WD and filling its 
Roche lobe. We can calculate at which velocity should it have been ejected.

\noindent
We would have, for the orbital motion of the star around the center of mass
of the binary:

\begin{equation}
v^{2}_{\rm orb} = {GM_{WD}\over a(1 + q)} 
\end{equation}  

\noindent
where $a$ is the orbital separation and $q \equiv M_{\rm comp}/M_{WD}$. In our 
case we have $M_{WD} = 1.4 M_{\odot}$ and $M_{\rm comp} = 0.44 M_{\odot}$, 
corresponding to a M1 dwarf (Pecault \& Mamajek 2013). So, in our case, 
$q$  = 0.314. 
The radius of a M1 dwarf is $R_{M} = 0.446 R_{\odot}$ (same source). 

\noindent
We have, on the other hand, the Eggleton (1983) approximate formula for the
Roche lobe radius $R_{L}$ of the secondary star in a binary:

\begin{equation}
R_{L} = a\left[{0.49 \over 0.6 q^{-2/3} + {\rm ln}(1 + q^{1/3})}\right]
\end{equation}

We thus have for the orbital velocity, by making $R_{L} = R_{M}$ in (2) to 
obtain $a$ and substituting it in (1):
 
\begin{equation}
v^{2}_{\rm orbit} = {0.49\ GM_{WD}\over R_{M}(1 +  q)[0.6 q^{-2/3} + 
{\rm ln}(1 + q^{1/3})]}
\end{equation}

\noindent
and then, rounding to unity, $v_{\rm orbit}$ = 350 ~\kms. We have measured a 
total velocity $v_{tot} = 256^{+181}_{-70}$ ~\kms for
star  M\RomanNumeralCaps{5}-G272 (vector sum of tangential and radial 
velocities).  

\smallskip
\noindent
The orbital velocity 
is just an upper limit to the actual one before explosion, since some mass 
should have been stripped by the impact of the SN ejecta and the pre-explosion
radius and orbital separation should also have been larger, then.

\smallskip
\noindent
 Concerning the rotational velocity, 
 even before explosion rotation might have 
been slowed down due to angular momentum loss from the mass transfer to the 
WD. The collision with the ejecta of the SN can drastically reduce the 
rotational velocity. This  has been shown by Liu et al. (2013b) and Pan et 
al. (2014). In the 3D hydrodynamical simulations of Liu et al. (2013b), the 
rotational velocity of the companion is reduced to only 14\% to 32\% of 
its pre-explosion value. Similar results are quoted by Pan et al. (2014), 
with references to their previous work (Pan et al. 2012b, 2013).    
An extra mechanism to slow down the rotation of the companion star after the
impact of the SN ejecta would act during the evaporation phase of the surface
layers of the star (those that have not been ablated by the impact but have
absorbed enough energy to become unbound). If the wind remains tied to the
surface of the star by the magnetic field and is only lost at significant
distances above the surface, it will carry a lot of angular momentum, thus
significantly reducing the rotational velocity, (this
idea is being explored by X. Meng et al. 2023, in preparation).

Finally, it must be remembered that  what is actually measured is 
$v_{rot}$ sin $i$, where $i$ is the angle made by the rotation axis with the 
line of sight.

\noindent
Pan et al. (2012a) have calculated the amount of contamination by
Fe and Ni of the surfaces of SNeIa companions. They obtained
$\sim 10^{-5} M_{\odot}$ for MS star companions, $\sim 10^{-4} M_{\odot}$ for He
star companions, and $\sim 10^{-8} M_{\odot}$ for RG companions (see also Pan
et al. 2014). The observed contamination would, however, depend on the
degree of dilution of the contaminants with the stellar envelope.
Even in an early  M dwarf, most of the mass, from the surface down to close to 
the central layers, is convective. We should thus expect a strong dilution 
of the material captured from the ejecta, and thus only moderate overabundances
of Fe--peak elements.

    \section{HR diagrams}

    From the {\it Gaia} photometry and parallaxes, we can now construct 
    HR diagrams for our sample of stars. They can then be compared
    with 
    model predictions for different types of possible survivors from SNeIa
    explosions. There are only a few theoretical calculations of the evolution
    of SNeIa companions after being hit by te SN ejecta. Podsiadlowski
    (2003) modeled the evolution of a SG star of 2.1 $M_{\odot}$ for up to
    10,000 yr 
    after the explosion, and later Shappee et al. (2013) did the same for a
    MS companion of 1 $M_{\odot}$. In both cases, mass stripping from the
    impact was modeled as a fast wind and energy injection to the 
    layers of the companion that remained bound was parameterized. The
    results, for the luminosities and effective temperatures of the
    companions, 10,000 yr after the explosion, very widely differ, as it can
    be seen by comparing Fig. 1 in Podsiadlowski (2003) with Fig. 4 in
    Shappee et al. (2013). 
    
\noindent
    Di Stefano et al. (2011) (see also Justham 2011) have calculated the 
    evolution of SNIa companions for the case in which there is long enough
    delay between the end of mass transfer and the explosion (due to the fast
    rotation of the WD) to allow the companion to become a second WD 
    before the explosion takes place. Their calculations, however, stop 
    at this point and there are no existing hydrodynamical simulations of the 
    collision of the SN material with the WD, a prerequisite to know its
    state, thousands of years after the explosion.  
 
\noindent
    Meng \& Li (2019) (see also Meng \& Luo 2021)
    have calculated the luminosities and colors of possible MS companions
    ending as subdwarf (sdB) stars at the time of the SN explosion but, again, 
    neither the effects of the impact on them nor their subsequent evolution
    are included there. Bauer et al. (2019) have modelled the 
    collision of SNIa ejecta with a sdB star and also the evolution after
    the impact, providing various useful observational predictions.
    Pan et al.(2014), on the other hand, 
     start from 3D hydrodynamic models of the companions after being hit
    by the SN ejecta and follow their evolution hydrodynamically until
    hydrostatic (but not thermal) equilibrium is recovered. Those 3D models
    are then projected into 1D models, whose subsequent evolution is
    calculated using the {\it MESA} ({\it Modules for Experiments in Stellar 
    Astrophysics})
    code\footnote{docs.mesastar.org/en/release-r22.05.1}.
     In this way they predict the evolutionary tracks
    for MS companions up to 9,000 yr after being hit by the SN ejecta and
    those for He star companions until 1,000 yr after the explosion.

\noindent
    In Figure 13 we show the tracks followed by MS and He star companions
    with several masses and pre--explosion orbital parameters, taken from
    Pan et al. (2014). Model charateristics of MS companions are given in
    Table 4 (which reproduces part of their Table 1). The tracks are plotted
    on the
    $g$ vs $g - r$ plane (Sloan colors), assuming a distance of 2 kpc. The
    positions of the sdB stars from Meng \& Li (2019) are equally shown.
    Comparison is made with the stars in our {\it Gaia} sample and also
    with
    the larger sample from the {\it DECaPS} survey, which includes 
    stars within a
    wider range of distances. Both samples are corrected for reddening (see 
    Section 5, above). 

\noindent
    We clearly see that no sdB stars are present in any of the two samples.
     As for the He star companions, although a few stars from the 
    {\it DECaPS} survey
    lie not far from the ends of the tracks, this is consistent 
    with the
    dispersion due to the lack of distance boundaries there.
    MS companions are not close to our {\it Gaia} sample, either.
    They will be discussed next.

  \begin{table}[h!]
   \centering
      \caption{The main--sequence SN Ia companion 
       models of Pan et al. (2014)}
     \begin{tabular}{lcccccc}
      \tableline
      \tableline
     Model$^{1}$ & & $M_{b}$ & $R_{b}$ & $M_{a}$ & $R_{a}$ & $v_{linear}$ \\
     & & ($M_{\odot}$) & ($R_{\odot}$) & ($M_{\odot}$) & ($R_{\odot}$) & (km 
     s$^{-1}$) \\
     \tableline
  A & $\dots$ & 1.88 & 1.25 & 1.64 & 3.87 & 179 \\ 
  B & $\dots$ & 1.82 & 1.50 & 1.65 & 4.76 & 179 \\
  C & $\dots$ & 1.82 & 2.63 & 1.56 & 7.61 & 136 \\
  D & $\dots$ & 1.63 & 1.19 & 1.43 & 3.42 & 188 \\
  E & $\dots$ & 1.59 & 1.42 & 1.44 & 3.91 & 191 \\
  F & $\dots$ & 1.55 & 1.97 & 1.30 & 4.09 & 143 \\
  G & $\dots$ & 1.17 & 0.79 & 0.93 & 4.45 & 271 \\ 
      \tableline
    
     \end{tabular}

     $^{1}$  Subscript $b$ indicates before and subscript $a$ after\\  
      the SNe Ia explosion.
           
  \end{table}

\noindent
    In Figure 14 we compare, in the HR diagram log($L/L_{\odot}$) $vs$ log
    $T_{eff}$, the evolutionary tracks for the MS models A--G in Table 4 with
    our {\it Gaia} sample. 
    
\noindent
    The tracks calculated by Pan et al. (2014), as mentioned above, follow the 
    evolution of the
    possible companions up to 9,000 yr after the explosion. That covers, at 
    least in part, the estimated range of ages for the SNR.
    Only very few stars lie close to the ends of the evolutionary tracks
    A--G, and they do not display any kinematical peculiarity.

    \smallskip
    
   \noindent 
    Star  M\RomanNumeralCaps{5}-G272 lies quite apart from the tracks,
    but its mass 
    should be $M \simeq 0.44-0.50 M_{\odot}$, less than half the smaller mass in
    Table 4 and Figures 13 and 14 (that of model G).
    No simulations of the impact of the SN ejecta nor calculations of the
    subsequent evolution are available for stars of such a small mass.

\smallskip  

\noindent
     Very recently, Rau \& Pan (2022) have extended the calculations of the 
    post-impact evolution of MS companions down to 0.8 $M_{\odot }$, the result
    depending of the ratio of the orbital separation to the radius of the star
    at the time of the explosion, for a fixed explosion energy. They  
    follow the evolution up to more than 10$^{5}$ yr after the explosion. They 
    find that, for the lowest mass considered (0.8 $M_{\odot}$), the 
    luminostiy remains constant and can be as low as $\sim 2 L_{\odot}$ from 
    shortly after the explosion until $\sim 10^{3}$ yr later. From this
    point on, the luminosity starts 
    decreasing. A constant luminosity stage is common to all the 
    masses studied (0.8, 1, 1.5 and 2 $M_{\odot}$), such luminosity decreasing 
    fast when going from the larger to the lower masses (see their Fig. 2). 
    These new results appear consistent with an origin of star
    M\RomanNumeralCaps{5}-G272 from the impact of SNIa ejecta on a MS 
    companion of low mass.    

\smallskip 

\noindent  
    The simplest hypothesis for the origin would be that  
    M\RomanNumeralCaps{5}-G272 was a more massive MS star at the beginning of 
    mass transfer to the WD. The mass was reduced, by the process 
    of mass transfer to the WD, to a value close to the present one. There is
    little 
    mass stripping and energy input in the explosion, due to the compactness 
    of the star.
    Another possibility might be that  M\RomanNumeralCaps{5}-G272 was quite
    more massive than M\RomanNumeralCaps{5}-G272 even at the time of the 
    explosion, a sizeable fraction of its 
    mass having been stripped by the impact of the ejecta. It is what happens, 
    in variable extent, to
    the model stars in Table 4, although none of them ends having a mass as low
    as that of a M1-M2 dwarf, they being too massive for that, initially.
    A third possibility is that  M\RomanNumeralCaps{5}-G272 were an
    M dwarf already,
    at the start of mass transfer. The WD should then have been quite massive.
    M dwarfs have never been favourites as possible SN Ia companions. They
    would transfer H--rich material to the mass--accreting WD at a slow rate,
    so giving rise to nova--like outbursts, these expelling most of the
    accreted mass. Wheeler (2012), however, has presented a model in which
    the combined magnetic fields of the WD and of the M star lock them
    together. A kind of {\it  magnetic bottle} would then form and channel
    the mass transfer, so the WD would be 
    accreting matter through a limited polar area. Accretion rates would be
    enhanced due to the luminosity of that hot spot acting on the M--dwarf
    end of the bottle, mixing inhibited by the magnetic field, and the
    accreted material kept hot, thus avoiding thermonuclear runaway
    outbursts. Not being spun-up by the accretion, the WD  would be slowly 
    rotating. Given the high numbers of M dwarfs, this mechanism might  
    contribute to the observed SNe Ia rates.
    A fourth and last possiblity is, of course, that star
     M\RomanNumeralCaps{5}-G272 were unrelated to the SNR G272.2--3.2,
    its high
    velocity being due to past interactions with other stars, but
    then its path within the remnant would be a most extraordinary
    coincidence.
     We have examined several possibilities. Explanations for high-velocity
    stars include disruption of a close binary by the supermassive black 
    hole at the center of the Galaxy, with capture of a member and ejection 
    of the other; similar three-body interaction involving a black hole at
    the center of a globular cluster; tidal shredding of dwarf galaxies or 
    ejection from a nearby galaxy. None of these mechanisms is likely to 
    impart high
    velocity along the Galactic plane. Dynamical interaction between groups
    of massive stars leading to binary disruption has also been proposed, and 
    that would happen in the disk, but no low-mass 
    stars like  M\RomanNumeralCaps{5}-G272 would be ejected. There is no 
    stellar stream towards the site of the SNR. There is no pattern of 
    ejected stars from a globular cluster inside the SNR and its position 
    is far away from the supermassive black hole at the center of the 
    Galaxy.

    \begin{figure}[h!]
      \centering
      \includegraphics[width=0.65\textwidth]{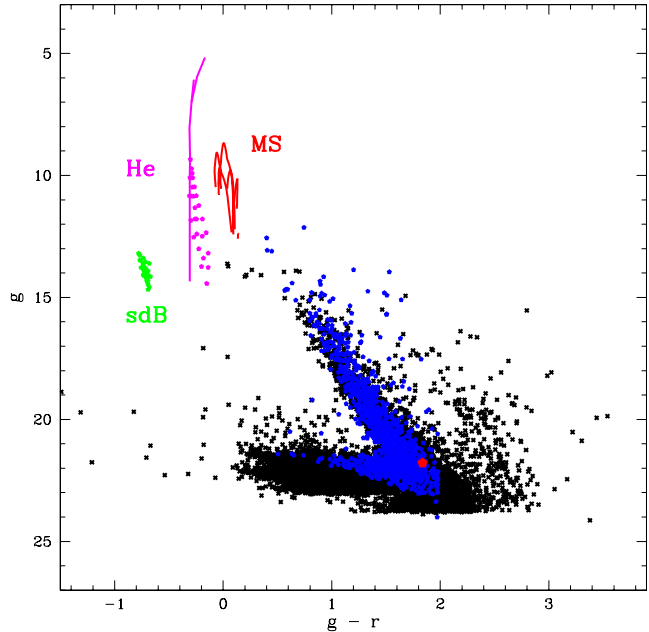}
      \vskip -2.5 true cm 
      \caption{$g$ $vs$ $g - r$ magnitudes, at the distance of the SNR
        G272.2-3.2 (taken here as 2 kpc), of the post--explosion evolutionary
        tracks of MS (red) and He (magenta) star companions (from Pan et al.
        2014a) and location of possible sdB companions (green) (from Meng \& Li
        2019),
        compared with our sample of stars (blue filled pentagons) from {\it
          Gaia}
      EDR3 and with the larger sample from the {\it DECaPS} survey (black 
      crosses), covering
      the same area of the sky but with no constraints on distance there. 
      The stars have been dereddened as discussed in Section 5. The red 
      dot marks the position of star  M\RomanNumeralCaps{5}-G272. Due 
      to the scale of the plot here, details of
      the MS evolutionary tracks are shown in Figure 14 only.}
\label{Figure 11}    
    \end{figure}

\begin{figure}[h!]
  \centering
   \includegraphics[width=0.75\textwidth]{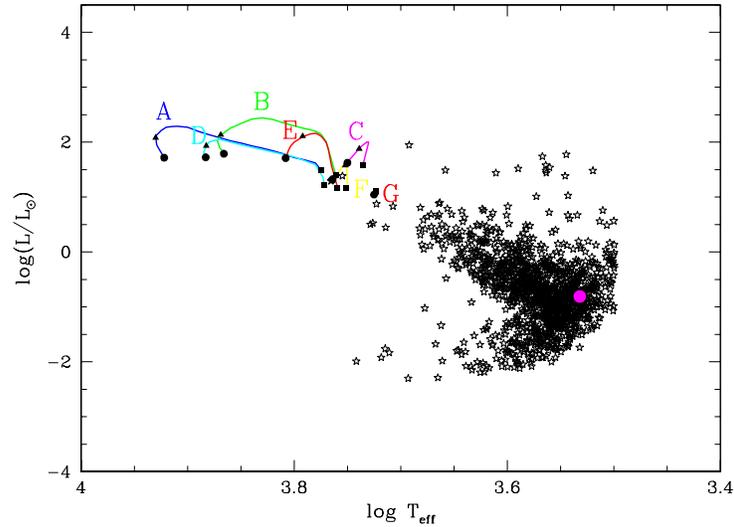}
   \vskip -3.5 true cm
  \caption{The HR diagram  of the stars of our sample, compared with the
    theoretical evolutionary paths of Pan et al. (2014a) for main--sequence
    star
    companions of SNe Ia after the explosion. The evolutionary
    tracks cover from the time the SN Ia companions recover hydrostatic
    equilibrium after being impacted by the SN ejecta to 9,000 years later.
    The 100, 500, 3,000 and 9,000 yr post--explosion stages are marked by
    filled squares, stars, triangles and circles, respectively. Star
    M\RomanNumeralCaps{5}-G272 is marked by a magenta dot. The stars
     have been dereddened as discussed in Section 5.}
  \label{Figure 12}
\end{figure}

\section{\bf Other high-proper motion stars}

\begin{figure}[ht!]
  \centering
  \includegraphics[width=0.45\textwidth]{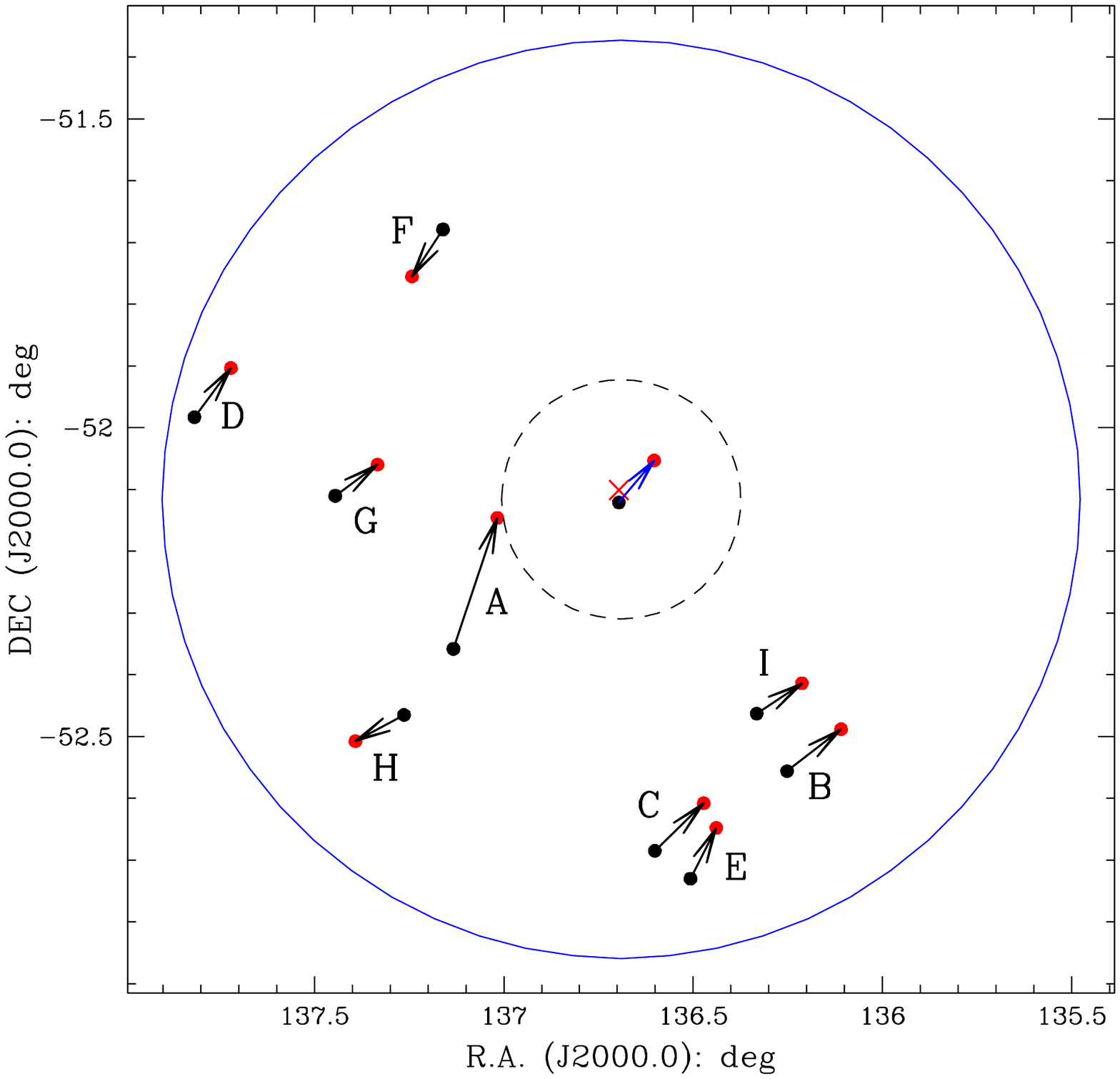}
  \caption{ The 9 stars (labelled A-I),  with total proper 
    motions higher than that 
    of M\RomanNumeralCaps{5}-G272 and located between 11 arcmin and 42.2 
    arcmin from the centroid of the 272.2-3.2 SNR (marked with a red cross).
    The dashed circle corresponds to the 11 arcmin radius. Present positions
    are marked with red dots and those 8,000 yr ago with black ones. 
    The motion of star M\RomanNumeralCaps{5}-G272 is shown by a blue arrow. We
    clearly see that none of these high-velocity stars can come from inside the
    SNR (approximately limited by a 9 arcmin radius).
    Positions and proper motions from {\it Gaia} EDR3 are very precise, so 
    variations of the trajectories within the error limits would hardly be 
    seen in the Figure.
}
  \label{Figure 16}
  \end{figure}

 Although our initial exploration of the SNR G272.2-3.2 has produced a good 
candidate to be the surviving companion of the exploding WD that produced 
the remnant, one may wonder whether a more extended search would produce some 
other candidate, moving faster than M\RomanNumeralCaps{5}-G272, as might be 
the case of some He star companions or of hypervelocity stars (produced by the 
D$^{6}$ mechanism mentioned in the Introduction). 
We will thus further explore the region around G272.2-3.2, in order to check 
for such extra possibilities.

\smallskip
\noindent  
 In Section 2 and thereafter, a search radius of 11 arcmin around the 
centroid of SNR G272.2-3.2 has been adopted for the exploration. We remind 
that this radius corresponds to the angular distance covered by a star at 2 
kpc from us, moving at 500 \kms for 12,000 yr, perpendicularly to the line of 
sight. One might now take the lower
limit to the distance to G271.2-3.2 (1 kpc), the upper limit to its age (12,000 
yr) again and include stars with tangential velocities of up to 1,000 ~\kmso 
(possible He star companions): a search radius of 42.2 arcmin results 
(almost four times that used for our initial search). We have also made such
extended exploration. In Figure 15 we show the present and 
past positions (8,000 yr ago) of the 9 stars with total proper motions larger 
than that of star M\RomanNumeralCaps{5}-G272, in the new area.  We clearly 
see, from the Figure, that none of the 9 stars (whose characteristics 
are listed in Table 5) can have anything to do with the SNR. Stars 
with smaller proper motions (but still at more than 3$\sigma$ above the
average), located within the new ring, are not shown in Figure 15, but it 
is obvious that, they being located outside the 11 arcmin radius around the 
centroid at present, and moving more slowly across the sky than 
M\RomanNumeralCaps{5}-G272, they can hardly even have been inside the area 
now covered by the remnant (which has an approximate radius of 9 arcmin only), 
8,000 yr ago.

   \begin{table*}
      \centering
      \caption{
        The 9 stars with total proper motions larger than that of 
        M\RomanNumeralCaps{5}-G272, within the 42.2 arcmin radius from 
        the centroid of G272.2-3.2.}
      \begin{tabular}{lcccccc}
   \tableline  
    \tableline
Star & RA   & DEC & $d$  & $\mu^{*}_{\alpha}$ & $\mu_{\delta}$ & $v_{tan}$ \\
     & (deg)& (deg) & (kpc) &(mas yr$^{-1}$) & (mas yr$^{-1}$)& (\kmso) \\ 
\tableline
A  & 137.017 & -52.146 & 2.11$^{+ *}_{-1.31}$    & -25.641 & 76.381 & 
807$^{+ *}_{-501}$ \\  
B  & 136.109 & -52.488 & 1.30$^{+0.65}_{-0.32}$  & -39.558 & 30.548 & 
308$^{+155}_{-75}$ \\
C  & 136.472 & -52.608 & 1.00$^{+0.36}_{-0.20}$ & -35.620 & 34.783 & 
236$^{+85}_{-47}$ \\
D  & 137.720 & -51.903 & 2.96$^{+ *}_{-1.94}$ & -26.766 & 35.898 & 
629$^{+ *}_{-412}$ \\
E  & 136.439 & -52.648 & 1.87$^{+1.76}_{-0.61}$  & -18.969 & 37.134 & 
370$^{+349}_{-120}$ \\
F  & 137.242 & -51.755 & 1.19$^{+11.48}_{-0.56}$  &  22.633 &-34.213 & 
232$^{+2236}_{-109}$ \\
G  & 137.333 & -52.060 & 1.43$^{+2.40}_{-0.55}$  & -33.587 & 22.790 & 
276$^{+462}_{-106}$ \\
H  & 137.392 & -52.507 & 1.03$^{+1.54}_{-0.38}$  &  35.488 &-19.089 & 
197$^{+294}_{-73}$ \\
I  & 136.212 & -52.414 & 2.57$^{+ *}_{-1.83}$  & -32.943 & 22.413 & 
486$^{+ *}_{-346}$ \\
\tableline
\end{tabular}
\end{table*}

\smallskip
\noindent
 Concerning the high-proper motion stars in the extended 42.2 arcmin 
radius area, 
it must be noted that the sample comprises 54,035 stars. An estimated fraction
of 0.002 of the stars in the Solar neighbourhood belong to the halo population
(see Konishi et al. 2015, for instance).
Thus, the sample should include $\sim$ 100 of them, with velocities of at least
220 ~\kmso relative to the Local Standard of Rest (Du et al. 2018). From 
its metallicity we know that M\RomanNumeralCaps{5}-G272 is not one of  
these.
 
\smallskip

\noindent
A possible scenario for the production of SNe Ia is, as we said in the
Introduction, the {\it dynamically driven double--degenerate,
  double--detonation} scenario (D$^{6}$) (Shen \& Moore 2014; Shen \& Schwab
 2017; Shen et al. 2018). In this scenario, detonation of a He layer at the 
 surface of the more massive,
mass--accreting WD induces, by compression, its whole detonation. That would
happen when no much mass has yet been transferred
from the less massive WD, which has not been tidally disrupted. It
would, therefore, be ejected from the system at its orbital velocity. The
system being extremely compact, the velocity should be very high ($>$ 1,000
~\kmso). In fact, three objects have been found, in the {\it Gaia} EDR2, moving
at 1,000--3,000 ~\kmso, which might be former WD companions in a pre--SN system
(Shen et al. 2018).

\noindent
Since SNR G272.2--3.2 is comparatively young, even a
possible hypervelocity former companion of the SN cannot have traveled very
far. Even moving at 3,000 ~\kmso, perpendicularly to the line of sight, for
12,000 yr, its trajectory on the sky, assuming a distance of 2 kpc, would only
reach $0.87^{o}$ away from the site of the explosion. Thus, exploring the
region up to a full degree from the centroid of the SNR is enough to catch any
possible hypervelocity object produced by the explosion.

\noindent
 This new exploration has produced a sample of 112,704 stars, none of them 
with a
total proper motion larger than that of star A in Table 5 and Figure 15. All 
the new high-velocity stars are outside the  previously explored 11 arcmin
radius, so they can in no way have originated from the SNR.

\noindent
 To be exhaustive, one can take a distance of 1 kpc, with the same 
tangential velocity
of 3,000 \kmso. That gives a radius of 2.1 degrees. The sample then comprises
514,072 stars, and again there is no star with total proper motion 
surpassing that of star A.

\noindent
 From a {\it Goodman} spectrum, star A 
({\it Gaia} EDR3 5323871314998012928) is a MV1-M2V star, with solar 
metallicity. Although, due to the direction of its motion (see Figure 15), it 
can not have originated from inside the SNR G272.2-3.2, its tangential 
velocity, if it were at a distance of 2.11 kpc (see Table 5), would be of 
$\sim$ 807 ~\kms (with a large error propagated from that on the parallax).
 From a 
{\it MIKE} spectrum (see 
Figure 16) we have measured its radial velocity, which is only of $v_{helio} 
= 57.5\pm$0.4 ~\kms or $v_{LSR} = 42.2\pm$0.4 ~\kms. We thus think that 
star A is at a distance close to its lower limit in Table 5, which places
its total velocity well below the range of the hypervelocity stars.

\begin{figure*}
\centering
\includegraphics[width=0.65\textwidth]{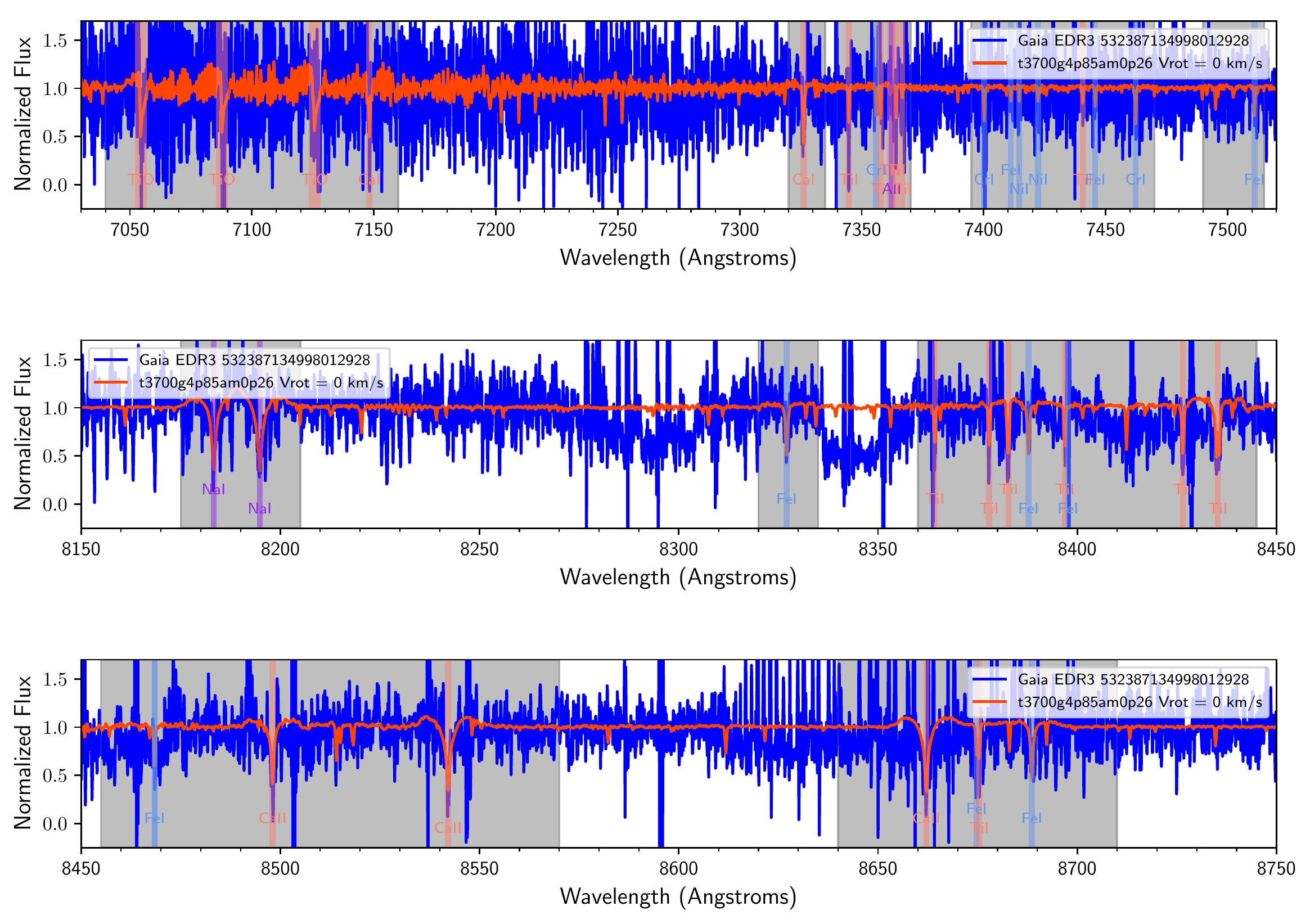}
\includegraphics[width=0.65\textwidth]{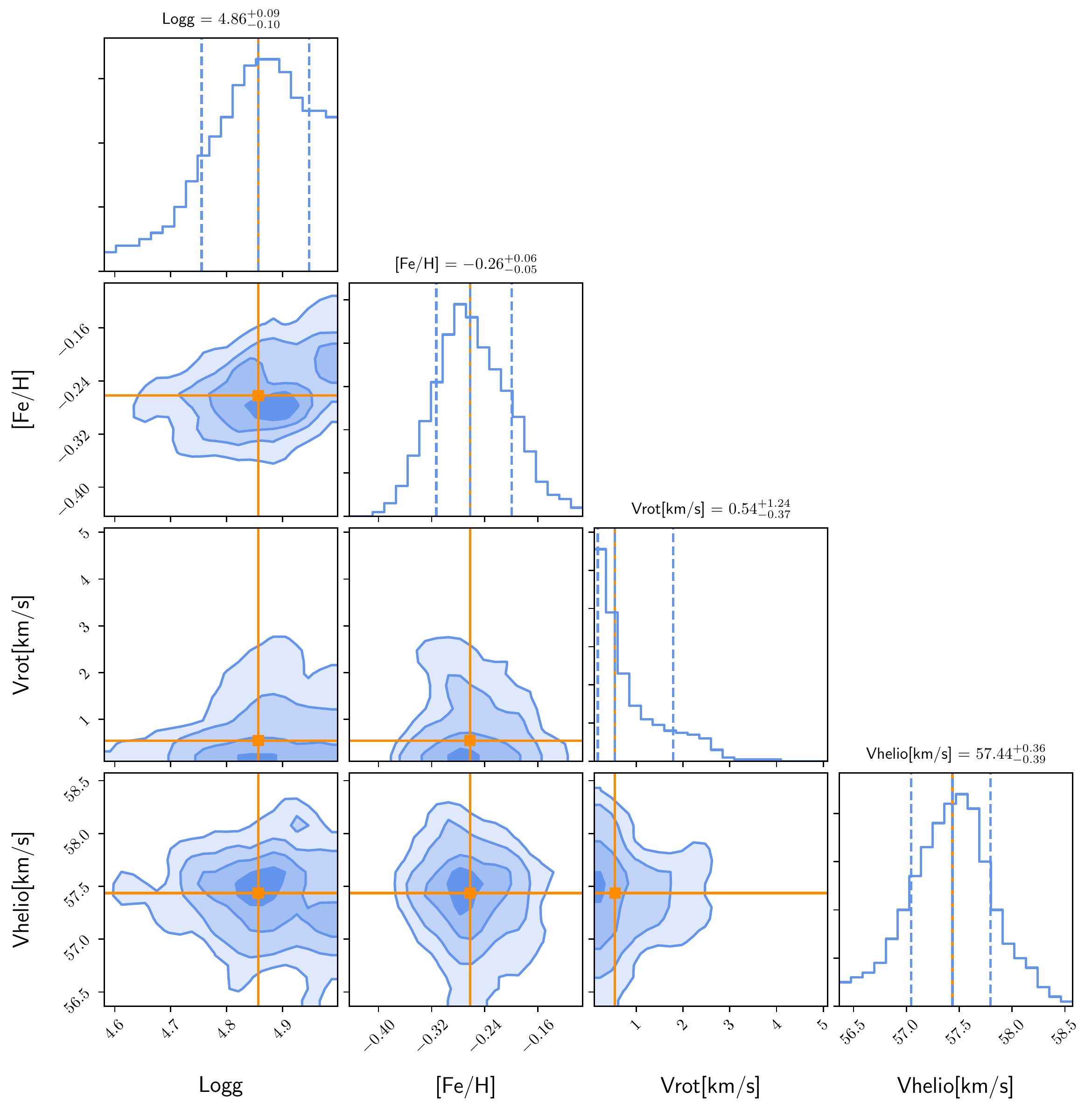}
\caption{ {\it MIKE} spectrum of {\it Gaia} EDR3 5323871314998012928
    (upper panel) and  posterior 
distributions of stellar parameters and metallicity (lower panel).} 
\label{Figure 7}
\end{figure*} 

\noindent
A relevant question is whether hypervelocity stars like those identified
by Shen et al. (2018) would be detectable at the distance of the SNR
G272.2--3.2. Two of them, labelled D6--1 and D6--3 in their Table 1, are at
distances of 2.1--2.3 kpc and have {\it Gaia} magnitudes $G$ of 17.4--18.3, so
they would have clearly been seen in our survey. As for D6--2 (LP 398-9), it 
is at a distance $d = 0.84\pm0.04$ kpc only (from
this last reference), and has $G$ = 16.97 mag. So, relocated at 2 kpc (the
mean distance of our survey), it would still have $G$ = 18.85 mag and be also
detected. In addition, it has been associated with a $\sim$ 10$^{5}$ yr old
SNR. And the evolution of a WD after being impacted, heated and bloated by the
SN ejecta, should be cooling, contracting and fading, so at only 12,000 yr (the
upper limit for the age of G272.2--3.2), such an objet should be more luminous
than D6--2/LP 398--9, if anything.

\noindent
Therefore, we can quite confidently conclude that no hypervelocity star of 
the type D$^{6}$ (Shen et al. 2028) has been produced by the explosion giving 
rise to G272.2--3.2.

\section{Summary and conclusions}

G272.2-3.2 is the SNR of a relatively recent ($\sim$ 7,500 yr old) SNIa which
had been unexplored up to now in search of possible surviving companions of 
the SN, though being, by its distance and location in the Galaxy, accessible 
to observations at all wavelengths.

\noindent
We have first used the parallaxes, proper motions and photometry from the 
{\it Gaia} EDR3, to explore the region within a circle of 11 arcmin radius 
around the centroid of the SNR and within a distance 1 kpc $\leq d 
\leq$ 3 kpc. That produced a sample of 3,082 stars. The surveyed area is 
larger than the SNR and encloses it. We then looked for kinematical signatures
of a possible SNIa companion star. We also had the {\it Gaia} photometry, used 
in a subsequent step. We checked this photometry against that from the 
{\it DECaPS} survey and found complete agreement between them. 

\noindent
From the statistics of the proper motions of the stars in our sample, one of
them,  M\RomanNumeralCaps{5}-G272, appears as a clear outlier, with a 
total proper motion 8.9$\sigma$ above the mean. We checked this peculiarity
against the Besan\c con model of the Galaxy, which confirmed it. The peculiar
motion is mostly along the Galactic plane. Spectra obtained for this star 
have allowed us to measure its radial velocity as well. The total velocity is
$v_{tot} = 256^{+181}_{-70}$ ~\kms, which falls within the range of velocities 
expected for small mass companions of SNeIa. Reconstruction, from the proper 
motions, 
of the past trajectory shows that the star, which is now near the 
periphery of the SNR, was at the center, 6,000-8,000 yr 
ago. Given the long path traced by the star, this coincidence is most 
significant. Such trajectory is unique among the 3,082 stars of the sample.

\noindent
Spectra obtained with the {\it MIKE} spectrograph at the 6.5m {\it Clay} 
telescope and with the {\it Goodman} spectrograph at the 4.1m {\it SOAR} 
telescope, allowed the classification of  M\RomanNumeralCaps{5}-G272 
as a M1-M2 dwarf, with solar metallicity, by comparing them with
{\it BOSS} templates. They also showed that the 
extinction, in the direction of the SNR, was small, which is relevant for the 
photometry. The star thus has a mass $M$ = 0.44-0.50 $M_{\odot}$ and radius 
$R$ = 0.446-0.501 $R_{\odot}$. There is agreement between the 
measured total velocity of  M\RomanNumeralCaps{5}-G272 and the ejection velocity
of a M1-M2 dwarf in close orbit with a 1.4 $M_{\odot}$ WD, when the binary is 
disrupted by a SNIa explosion.

\noindent
The spectrum obtained with the {\it MIKE} spectrograph at the 6.5m {\it Clay} 
telescope has allowed to establish the values of the  stellar parameters of 
 M\RomanNumeralCaps{5}-G272 and also to make a chemical analysis of its 
 surface through  synthetic spectra analysis (see section 6). We have then 
 $T_{eff}$ = 3,800 K (the range can be 3,600-3,850 K, 
given the systematic uncertainties), log $g$ = 4.46 and 
metallicity about solar: [Fe/H] = -0.32 (with 0.3 dex in error). We also 
looked for signs of
overabundances of Fe-peak elements coming from the SN ejecta, but systematic
uncertainties preclude any conclusion. In any case, a M1-M2 dwarf being almost
fully convective, a strong dilution of this material should be expected.

\noindent
The {\it Gaia} EDR3 photometry, together with the estimate of the reddening 
and the knowledge of the distances, allows to construct the color-magnitude 
and the HR diagrams of the stars of our sample. They are compared with the 
existing models of the evolution of proposed SNIa companions. That has allowed
to discard the presence of RGs, He stars and sdB companion stars in our sample. 

\noindent
The models for MS companions are more luminous and hotter than the sampled
stars.  M\RomanNumeralCaps{5}-G272 is fainter and cooler than any 
of the model stars, but all the models are for much more massive stars, 
so  the comparison is not significant. 

\noindent
To be exhaustive, the possibility that the SN were produced through the 
$D^{6}$ mechanism has been checked by the exploration of a circle with 2.1 
degree radius around the SNR. None has been found.

\noindent
 We have examined all evolutionary paths that might have led to the SNIa 
that produced SNR G272.2-3.2.   

\noindent
We have a very kinematically peculiar star,  M\RomanNumeralCaps{5}-G272
with all signs of having been ejected by a explosion taking place at the 
center of a SNIa SNR. Such characteristics make it unique. Although 
its  being a late--type, small star could come as a surprise, the evidence in 
its favour is very solid, coming from its kinematics and its 
trajectory inside the SNR. 

\noindent
In conclusion, we have found a possible companion star of the SNIa that 
resulted in  SNR G272.2-3.2, 
with much evidence in its favor. This would be the case, therefore, of a SD 
scenario involving a M dwarf star. Since, from the chemical abundances in its 
ejecta, the explosion that gave rise to SNR G272.2-3.2 was a normal SNIa and 
M dwarfs are the most abundant stars in the Galaxy, that opens up the prospect
for many SNeIa to have the same origin.

\bigskip

\noindent
Based on observations obtained at the Southern Astrophysical Research (SOAR) 
telescope (NOIRLab Prop. 2022A-606104), which is a joint project of the 
Ministerio da Ciencia, Tecnolog\'{\i}a e Innovaciones (MCTI/LNA) do Brasil, 
the US National Science Foundation's NOIRLab, the University of North Carolina 
at Chapel Hill (UNC), and Michigan State University (MSU).
We are grateful to the SOAR staff for their help in performing the 
observations of this project. 

\noindent
This paper includes data gathered with the 6.5 meter Magellan Telescopes 
located at Las Campanas Observatory, Chile.

\noindent
This work has made extensive use of the {\it Gaia} EDR3.
{\it Gaia} data are being processed by
the {\it Gaia} Data Processing and Analysis Consortium (DPAC).
Funding for the DPAC is provided by national institutions, in
particular the institutions participating in the {\it Gaia} MultiLateral Agreement (MLA). The Gaia mission website is https:
//www.cosmos.esa.int/gaia. The {\it Gaia} archive website is
https://archives.esac.esa.int/gaia.
This work was (partially) funded by the Spanish
Ministry of Science and Innovation (MICINN), the Agencia Estatal de Investigación (AEI) 10.13039/501100011033 and by "ERDF A way of making Europe"
by the “European Union” through grants RTI2018-095076-B-C21 and PID2021-122842OB-C21, and the Institute of Cosmos Sciences University of Barcelona (ICCUB, Unidad de Excelencia ’Mar\'{\i}a de Maeztu’) through grant CEX2019-000918-M.

\noindent
PR-L also acknowledges support from grant PGC2018-095157-B-I00, from the
MICINN.
JIGH acknowledges financial
support from the MICINN 
PID2020-117493GB-I00, and also from the Spanish MICINN 2013 Ramon y Cajal 
program RYC-2013-14875.
R.C acknowledges financial
support from grant  PGC2018-095157-B-I00 from Spanish MICINN.
L.G. acknowledges financial support from the Spanish MICINN, AEI
10.13039/501100011033, and the European Social Fund (ESF) "Investing in your future" under the 2019 Ram\'on y Cajal program RYC2019-027683-I and the PID2020-115253GA-I00 HOSTFLOWS project, from Centro Superior de Investigaciones Cient\'ificas (CSIC) under the PIE project 20215AT016, and the program Unidad de Excelencia Mar\'ia de Maeztu CEX2020-001058-M. 

\noindent
PRL would like to thank Evan Bauer and Warren Brown 
at the Harvard-Smithsonian  Center for Astrophysics, and  Christian Knigge
from the University of Southampton, for conversations. The authors would
like to thank the anonymous referee for the valuable comments on the manuscript.

\bigskip
\bigskip
{\bf References}

\bigskip

Bauer, E.B., White, C.J., \& Bildsten, L. 2019, ApJ, 887, 68

Bedin, L.R., Ruiz--Lapuente, P., Gonz\'alez Hern\'andez, J.I., et al.
2014,  MNRAS, 439, 354

Bernstein, R., Shectman, S. A., Gunnels, S. M., et al. 2003, SPIE, 4841,
1694. doi:10.1117/12.461502 

Branch, D., \& Wheeler, J.C. 2017, {\it Supernova Explosions} 
(Berlin, Springer)

Castelli, F., \& Kurucz, R. L. 2003, Modelling of Stellar Atmospheres,
210, A20

Clemens, J.C., Crain, J.A., \& Anderson, R. 2004, SPIE, 5492, 455

Di Stefano, R., Voss, R., \& Claeys, J.S.W. 2011, ApJL, 738, L1

Du, C., Li, H., Liu, S., Donlon, T., \& Newberg, H.J. 2018, ApJ, 863, 87

Duncan, A.R., Stewart, R.T., Campbell-Wilson, D., Haynes, R.F., 
Aschenbach, R., \& Jones, K.L. 1997, MNRAS, 289, 77  

Edwards, Z.I., Pagnotta, A., \& Schaefer, B.E. 2012, ApJ, 747, L19

Eggleton, P.P. 1983, ApJ, 268, 368

Foreman-Mackey, D., Hogg, D. W., Lang, D., et al. 2013, PASP, 125, 306. 
doi:10.1086/670067

Gonz\'alez Hern\'andez, J.I., Ruiz-Lapuente, P., Filippenko, A.V., et al. 2009, ApJ, 691, 1

Gonz\'alez Hern\'andez, J.I., Ruiz--Lapuente, P., Tabernero, H.M., et
al. 2012, Natur, 489, 533

Greiner, J., \& Egger, R. 1993, IAU Circ. No. 5709

Greiner, J., Egger, R., \& Aschenbach, B. 1994, A\&A, 286, L35

Han, Z. 2008, ApjL, 677, L109 

Harrus, I.H., Slane, P.O., Smith, R.K., \& Hughes, J.P. 2001, ApJ, 552,
614

Hayden, B., Rubin, D., Boone, K., et al. 2021, ApJ, 912, 87

Houdashelt, M. L., Bell, R. A., Sweigart, A. V., \& Wing, R. F. 2000, AJ, 119, 1424

Iben, I.,Jr., \& Tutukov, A.V. 1984, ApJS, 54, 355

Jenkins, J.S., Ramsey, L.W., Jones, H.R.A., et al. 2009, ApJ, 704,975

Jordi, C., Gebran, M., Carrasco, J. M., et al. 2010, A\&A, 523, A48

Justham, S. 2011, ApJL, 730, L34

Kashi, A., \& Soker, N. 2011, MNRAS, 417, 1466

Kamitsukasa, F., Koyama, K., Nakajima, H., et al. 2016, PASJ, 685, 7

Kelson, D.D. 2003, PASP, 115, 688

Kelson, D.D., Illingworth, G.D., van Dorkum, P.G., \& Franx, M. 2000, 
ApJ, 531, 159

Kerzendorf, W.E., Schmidt, B.P., Asplund, M., et al. 2009, ApJ, 701, 1665

Kerzendorf, W.E., Schmidt, B.P., Laird, J.B., et al. 2012, ApJ, 759, 7

Kerzendorf, W.E., Yong, D., Schmidt, B.P., et al. 2013, ApJ, 774, 99

Kerzendorf, W.E., Childress, M., Scharw\"chter, J., Do, T., \& Schmidt,
B.P. 2014, ApJ, 782, 27

Kerzendorf, W.E., Long, K.S., Winkler, P.F., \& Do, T. 2018a, MNRAS,
479, 5696

Kerzendorf, W.E., Strampelli, G., Shen, K.J., et al. 2018b, MNRAS, 479,
192

Kesseli, A.Y., West, A.A., Veyette, M., et al. 207, ApJS, 230, 16

Kesseli, A.Y., West, A.A., Veyette, M., et al. 2020, Astrophysics Source 
Code Library, ascl:2002.011. PyHammer: Python spectral typing suite. 

Konishi, M., Hiroshi, S., Sumi, T., et al. 2015, PASJ, 67, 1

Kurucz, R. L., Furenlid, I., Brault, J., et al. 1984, National Solar 
Observatory Atlas, Sunspot, New Mexico: National Solar Observatory, 1984

Leahy, D.A., Ranasinghe, S., \& Gelowitz, M. 2020, ApJS, 248, 16

Li, C.-J., Chu, Y.-H., Gruendl, R.A., et al. 2017, ApJ, 836, 8

Li, C.-J., Kerzendorf, W.E., Chu, Y.-H., et al. 2019, ApJ, 886, 99

Litke, K.C., Chu, Y.-H., Holmes, A., et al. 2017, ApJ, 837, 111

Liu, Z.-W., Pakmor, R., Seitenzahl, I.R., et al. 2013a, ApJ, 774, 37

Liu, Z.-W., Pakmor, R, R\"opke, F.K., et al. 2013b, A\&A, 554, A109

Liu. Z.-W., R\"opke, F.K., \& Zeng, Y. 2022, ApJ, 928, 146

Livne, E. 1990, ApJL, 345, L53

Livne, E., \& Arnett, D. 1995, ApJ, 452, 62

Lopez, L.A., Ramirez-Ruiz, E., Hupperkothen, D., Badenes, C., \&
Pooley, D.A. 2011, ApJ, 732, 114

Marfil, E., Tabernero, H. M., Montes, D., et al. 2021, A\&A, 656, A162. 
doi:10.1051/0004-6361/202141980

Marietta, E., Burrows, A., \& Fryxell, B. 2000, ApJS, 128, 615

McEntaffer, R.L., Grieves, N., DeRoo, C.L., \& Brantseg, T. 2013, ApJ,
774, 120

Maoz, D., Mannucci, F., \& Nelemans, G. 2014, ARA\&A, 52, 107

McCutcheon, C., Zeng, Y., Liu, Z.-W., Izzard, R.G., Pan, K.C., 
Chen, H.-L., \& Han, Z. 2022, MNRAS, 514, 4078

Meng, X., \& Li, J. 2019, MNRAS, 482, 5651

Meng, X.-C., \& Luo, Y.-P. 2021, MNRAS, 507, 4603

Nomoto, K. 1982, ApJ, 253, 798

Pakmor, R., R\"opke, F.K., Weiss, A., \& Hillebrandt, W. 2008, A\&A,
489, 943

Pakmor, R., Kromer, M., Taubenberger, S., et al. 2012, ApJL, 747, L10

Pan, K.-C., Ricker, P.M., \& Taam, R.E. 2012a, ApJ, 750, 151

Pan, K.-C., Ricker, P.M., \& Taam, R.E. 2012b, ApJ, 760, 21

Pan, K.-C., Ricker, P.M., \& Taam, R.E. 2013, ApJ, 773, 49

Pan, K.-C., Ricker, P.M., \& Taam, R.E. 2014, ApJ, 792, 71

Passeger, V. M., Bello--Garc\'\i a, A., Ordieres--Mer\'e, J., et al. 2022,
A \& A, 658, A194

Pecaut, M.J., \& Mamajek, E.E. 2013, ApJS, 208, 9

Perlmutter, S., Aldering, G., Goldhaber, G., et al. 1999, ApJ, 517, 565

Podsiadlowski, P. 2003, arXiv:0303.660

Rau, S.-J., \& Pan, K.-C. 2022, ApJ, 933, 38

Reiners, A., Zechmeister, M., Caballero, J. A., et al. 2018, A\&A, 612,
A49. doi:10.1051/0004-6361/201732054

Riess, A..G., Filippenko, A.V., Challis, P., et al. 1998, AJ, 116, 1009

Rose, B.M.,Rubin, D., Cikota, A., et al. 2020, ApJL, 896, L4

Rosswog, S., Kasen, D., Guillochon,J., \& Ramirez-Ruiz, E. 2009, ApJL, 705, L128

Ruiz-Lapuente, P. 1997, Sci, 276, 1813

Ruiz--Lapuente, P. 2004, ApJ, 612, 357

Ruiz--Lapuente, P. 2014, NewAR, 62, 15

Ruiz--Lapuente, P. 2019, NewAR, 85, 101523

Ruiz--Lapuente, P., Comer\'on, F., M\'endez, J., et al. 2004, Nature, 
431, 1069

Ruiz--Lapuente, P., Damiani, F., Bedin, L., et al. 2018, ApJ, 862, 124

Ruiz--Lapuente, P., Gonz\'alez Hern\'andez, J.I., Mor, R., et
al. 2019, ApJ, 870, 135

Sezer,\& G\"ok, F. 2012, MNRAS, 421, 3538

S\'anchez-Ayaso, E., Combi, J.A., Bocchino, J.F., et al. 2013, A\&A, 552,
A52

Schaefer, B.E., \& Pagnotta, A. 2012, Natur, 481, 164  

Shappee, B.J., Kochanek, C.S., \& Stanek, K.Z. 2013, ApJ, 765, 150

Shen, K.J., \& Moore, K. 2014, ApJ, 797, 46

Shen, K.J., \& Schwab, J. 2017, ApJ, 834, 180

Shen, K.J., Boubert, D., G\"ansicke, B.T., et al. 2018, ApJ, 865, 15 

Shields, J.V., Kerzendorf, W., Hosek, Jr., M.W., et al. 2022, 
ApJL, 933, L31

Soker, N. 2013, in IAU Symp. 281, Binary Paths to Type Ia Supernova Explosins, ed. R. Di Stefano, M.Orio, \& M. Moe (Cambridge UK, Cambridge Univ. Press), 72.

Tabernero, H. M., Marfil, E., Montes, D., et al. 2021, Astrophysics
Source Code Library. ascl:2111.016. SterParSyn: Stellar atmospheric
parameters using the spectral synthesis method. 

Tabernero, H. M., Marfil, E., Montes, D., et al. 2022, A\&A, 657, A66. 
doi:10.1051/0004-6361/202141763

van Dokkum, P.G. 2001, PASP, 113, 1420

Wang, B., \& Han, Z. 2012, NewAR, 56, 122

Webbink, R.F. 1984, ApJ, 277, 355

Wheeler, J.C. 2012, ApJ, 758, 123 

Whelan, J., \& Iben, I.,Jr. 1973, ApJ, 186, 1007 

Xiang, Y., \& Jiang, Z. 2021, ApJ, 918, 24

Yamaguchi, H., Badenes, C., Petre, R., et al. 2014, ApJL, 785, L2

\clearpage

\appendix
\centerline{\bf MIKE Spectrum Fits}

\bigskip

\noindent
The analysis of the combined high-resolution MIKE spectrum (R$\sim$28,000) of the target Star MV-G272 is 
depicted in Fig. 17, where we have zoomed into several spectral regions from the whole spectral range shown in
Fig. 12. We also display the residuals the observed minus computed synthetic spectra (O-C) to compare 
the observations and the model. We see some remaining features corresponding to residuals coming
from sky subtraction and the telluric spectrum.

\bigskip

\begin{figure}[ht!]
\centering
\includegraphics[width=0.70\textwidth]{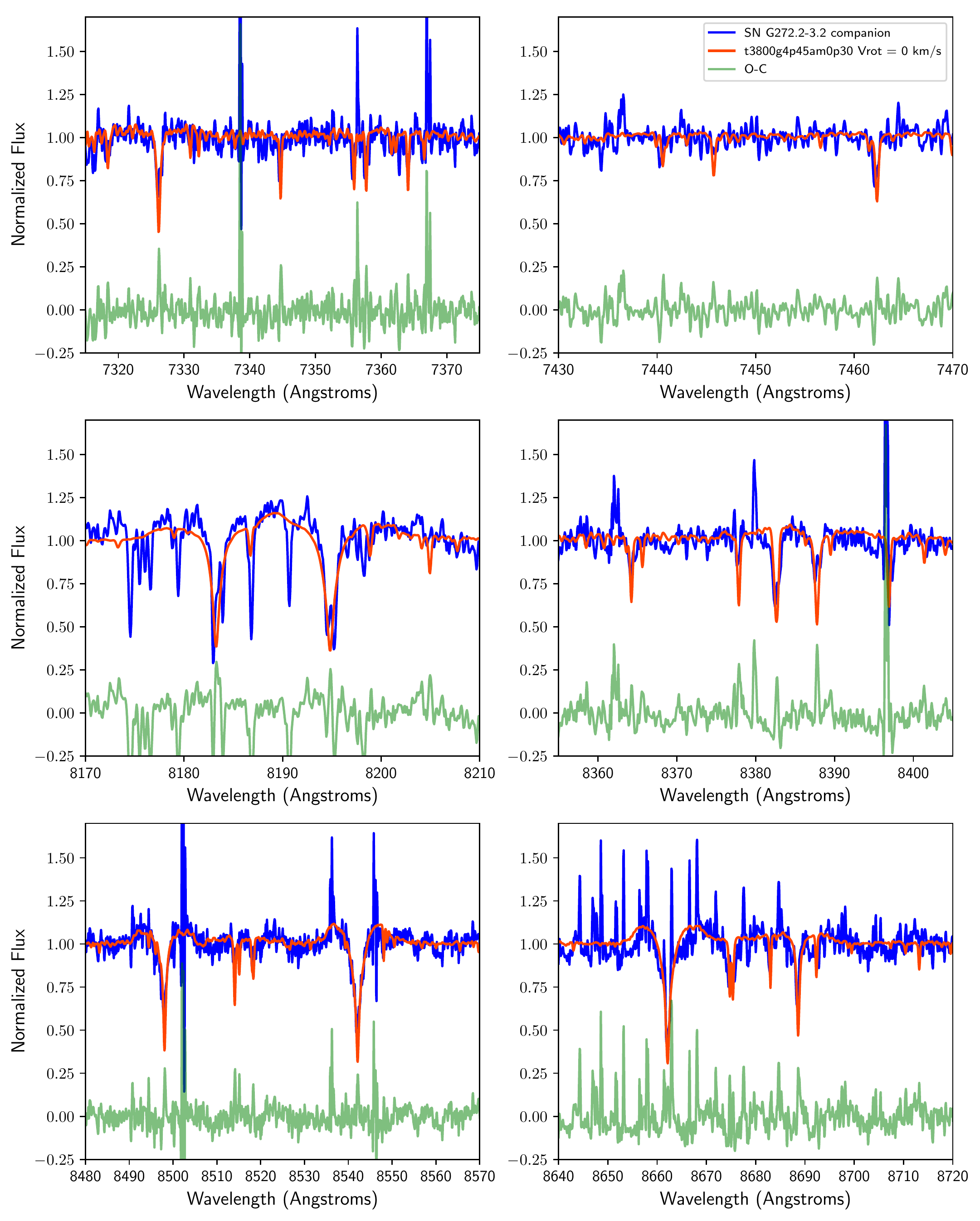}
\caption{Comparison of the observed {\it MIKE} spectrum of 
 M\RomanNumeralCaps{5}-G272 with the fitted spectrum, in the wavelength 
 ranges 7,320-7,370 \AA \ and 7,430-7,470 \AA \ (upper panel), 8,170-8,210 \AA
 \ and 8,360-8,400 \AA \  (middle panel) and
 8,480-8,570 \AA \ and 8,640-8,720 \AA \ (lower panel).}
\label{Figure 17}
\end{figure}

\noindent
We have tested our stellar parameter and metallicity determination using high-resolution CARMENES-VIS 
spectra (Reiners et al. 2018) with a resolution of R$\sim$94,600, degraded to a resolving power of R$\sim$28,000. We chose
two M1V stars Karmn J00183+440 (GX And) and Karmn J05415+534 (HD 233153), with
injected white noise down to S/N$\sim$10, as explained in Section 6.

\bigskip

\centerline{\bf A1. CARMENES Spectra}

\bigskip

\noindent
We display in Fig. 18 and 19, the resulting spectra of these two M dwarf stars
compare with a synthetic spectrum
in the same spectral range as the MIKE spectrum of the target Star MV-G272 display in Fig. 12.

s
\begin{figure}[ht!]
\centering
\includegraphics[width=0.85\textwidth,angle=0]{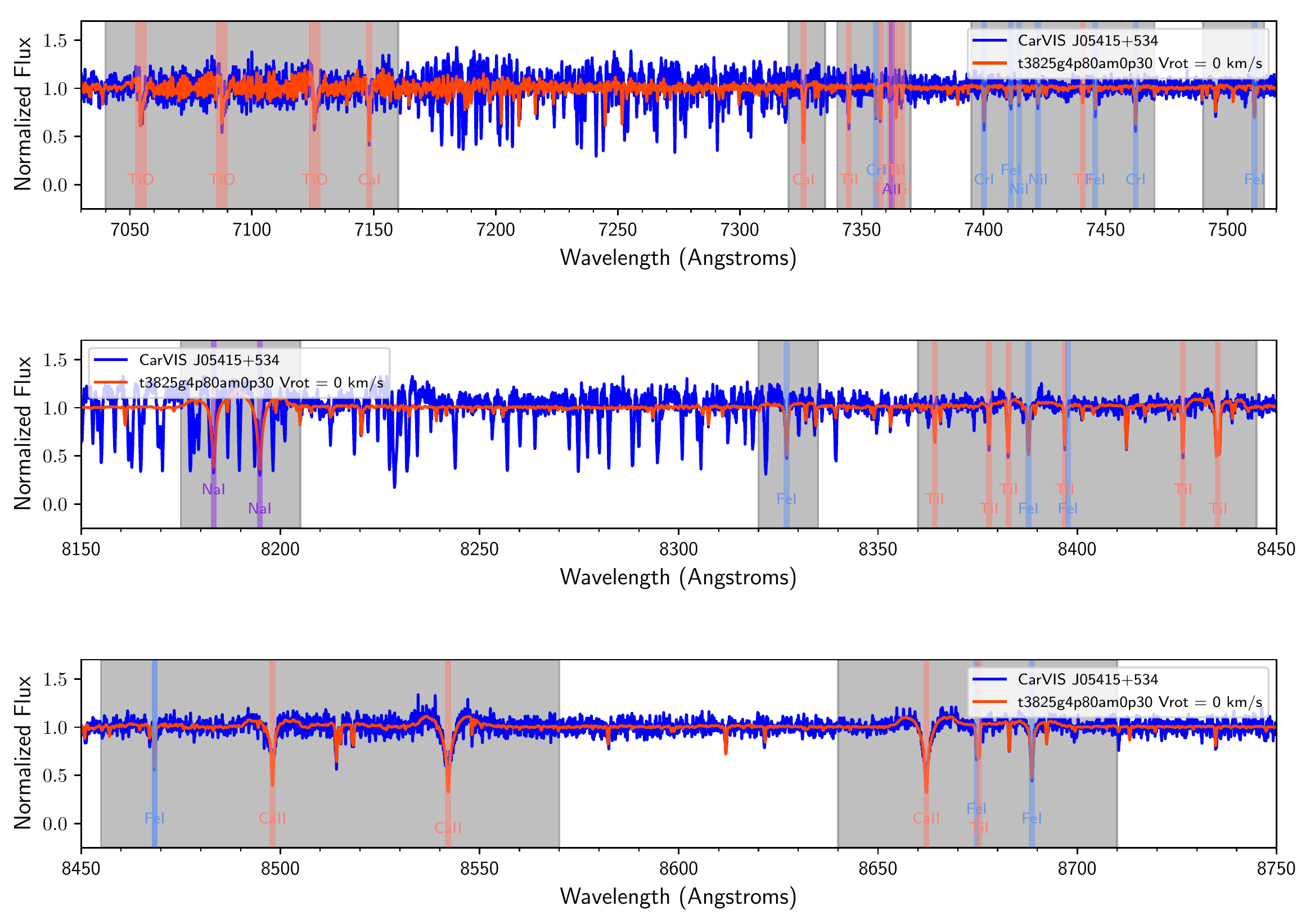}
\bigskip
\noindent
\begin{minipage}[b]{0.85\textwidth}
  \caption{Degraded and normalized CARMENES VIS 1D spectrum of star
    Karmn J05415+534 (HD 233153), 
corrected for barycentric radial velocity, degraded to a resolving power 
of $R\sim28,000$, with a signal-to-noise ratio of $\sim$~15 at 7400~{\AA}, and 
normalized to unity using a running mean filter with a width of 200 pixels 
at 0.069~{\AA} per pixel.
We also display an interpolated SYNPLE synthetic spectrum with the stellar 
parameters $T_{\rm eff}=3825$~K, $\log g =4.80$ and and metallicity 
[Fe/H]~$= -0.3$. 
The regions used to estimate the metallicity are shown in grey and the 
different lines 
used for chemical analysis are also highlighted.}
\end{minipage}
\label{figspec2}

\end{figure}

\begin{figure}[ht!]
\centering
\includegraphics[width=0.80\textwidth,angle=0]{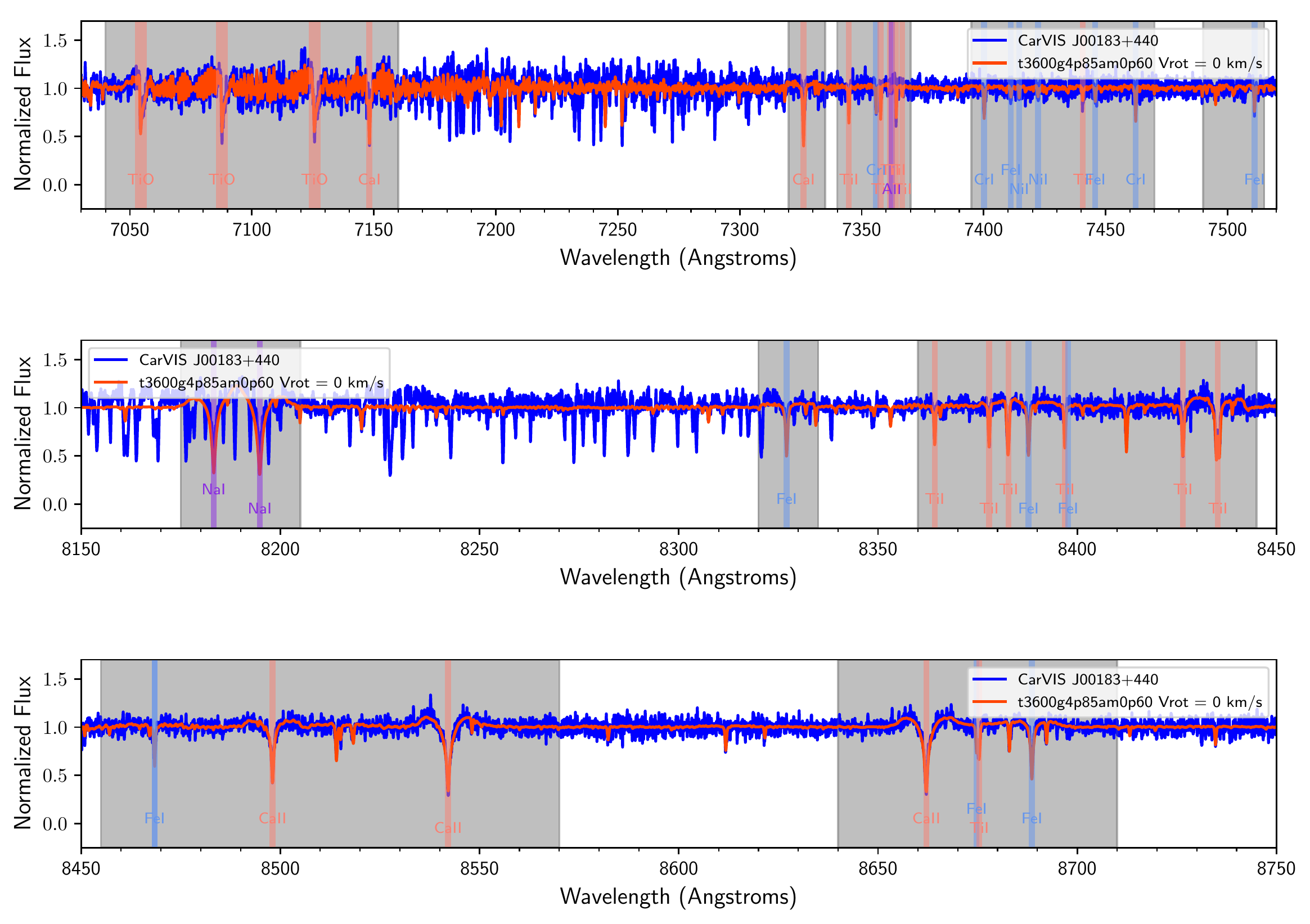}
\bigskip
\noindent
\begin{minipage}[b]{0.95\textwidth}
\caption{Normalized and degraded CARMENES VIS 1D spectrum of star Karmn 
J00183+440 (GX And), 
corrected for barycentric radial velocity, degraded to a resolving power 
of $R\sim28,000$, with a signal-to-noise ratio of $\sim$~15 at 7400~{\AA}, and 
normalized to unity using a running mean filter with a width of 200 pixels 
at 0.069~{\AA} per pixel. We also display an interpolated SYNPLE synthetic 
spectrum with the stellar parameters $T_{\rm eff}=3600$~K, $\log g =4.85$ and 
and metallicity [Fe/H]~$= -0.6$. The regions used to estimate the metallicity 
are shown in grey and the different lines used for chemical analysis are also 
highlighted.}
\end{minipage}
\label{figspec3}
\end{figure}

\end{document}